\documentclass[a4paper,11pt]{article}
\usepackage[utf8]{inputenc}
\usepackage{color}
\usepackage{amssymb}
\usepackage{amsmath}
\usepackage{comment}
\usepackage{graphicx}
\usepackage{epstopdf}
\usepackage{mathtools}
\usepackage{ragged2e}
\usepackage{hyperref}
\hypersetup{
    colorlinks=true,
    linkcolor=blue,
    filecolor=cyan, 
    urlcolor=red,
    citecolor=red,
} 
\usepackage[titletoc,toc,title]{appendix}
\numberwithin{equation}{section}
\usepackage{cleveref}
\usepackage[margin=2.5cm]{geometry}
\usepackage{cite}
\usepackage{epsfig}
\usepackage{float}
\usepackage{cancel}
\usepackage{braket}
\usepackage{physics}
\usepackage{amsfonts}
\usepackage{enumitem}
\usepackage[font={footnotesize,it}]{caption}
\usepackage{authblk}

\begin{document}
\begin{titlepage}

\title{Odd entanglement entropy in Galilean conformal field theories and flat holography}

\date{}

\author[1]{Jaydeep Kumar Basak\thanks{\noindent E-mail:~ jaydeep@iitk.ac.in}}
\author[1]{Himanshu Chourasiya\thanks{\noindent E-mail:~ chim@iitk.ac.in}}
\author[1]{Vinayak Raj\thanks{\noindent E-mail:~ vraj@iitk.ac.in}}
\author[1]{Gautam Sengupta\thanks{\noindent E-mail:~ sengupta@iitk.ac.in}}

\affil[1]{
Department of Physics\\

Indian Institute of Technology\\ 

Kanpur, 208016\\ 

India }

\maketitle
\begin{abstract}
\noindent
\justify

The odd entanglement entropy (OEE) for bipartite states in a class of $(1+1)$-dimensional Galilean
conformal field theories ($GCFT_{1+1}$) is obtained through an appropriate replica technique. In this context
our results are compared with the entanglement wedge cross section (EWCS) for $(2+1)$-dimensional asymptotically flat geometries dual to the $GCFT_{1+1}$ in the framework of flat holography. We find that our results are consistent with the duality of the difference between the odd entanglement entropy and the entanglement entropy of bipartite states, with the bulk EWCS for flat holographic scenarios.

\end{abstract}
\end{titlepage}
\tableofcontents
\pagebreak

\section{Introduction}
\label{sec1}
\justify

Characterization of quantum entanglement has emerged as a central issue for the investigation of diverse phenomena from condensed matter physics to issues of quantum gravity. In quantum information theory the entanglement entropy (EE) defined as the von Neumann entropy of the reduced density matrix appropriately characterizes the entanglement for bipartite pure states. Although this measure is relatively simple to compute for quantum systems with finite number of degrees of freedom it is usually intractable for extended quantum many body systems. Remarkably the entanglement entropy for bipartite states in $(1+1)$-dimensional conformal field theories ($CFT_{1+1}$s) could be obtained through a replica technique described in \cite{Calabrese:2004eu, Calabrese:2005in, Calabrese:2009qy}. For bipartite mixed states however the entanglement entropy receives contributions from irrelevant classical and quantum correlations and is hence an unsuitable entanglement measure for such states. Several mixed state entanglement and correlation measures have been proposed in quantum information theory although some of these involve optimization over local operations and classical communication (LOCC) protocols and are hence not easily computable. Some of these computable measures described in the literature include the entanglement negativity \cite{Vidal:2002zz, Plenio:2005cwa} and the reflected entropy \cite{Dutta:2019gen, Jeong:2019xdr}. Another novel computable measure for characterizing mixed state entanglement termed as the {\it odd entanglement entropy} (OEE) has been recently proposed in \cite{Tamaoka:2018ned}. The OEE may be loosely interpreted as the von Neumann entropy of the partially transposed reduced density matrix for the subsystem under consideration\footnote{\label{OEE-roughly-EE}The OEE is not exactly the von Neumann entropy  as the (partially transposed) density matrix utilized does not correspond to any physical state and may have negative eigenvalues.}. The authors in \cite{Tamaoka:2018ned} also obtained the OEE for the bipartite mixed state of two disjoint intervals in a $CFT_{1+1}$ through an appropriate replica technique. Furthermore it has been shown in \cite{Tamaoka:2018ned} that the holographic dual of the difference between the OEE and the EE is described by the bulk entanglement wedge cross section (EWCS) for the bipartite state in question\footnote{See \cite{Mollabashi:2020ifv, Kusuki:2019evw, Ghasemi:2021jiy, Angel-Ramelli:2020wfo, Berthiere:2020ihq, Dong:2021clv} for further developments on the OEE. }. It should also be noted here that the bulk EWCS has also been proposed as a holographic dual for several other correlation measures, for example the entanglement of purification \cite{Takayanagi:2017knl}, the reflected entropy \cite{Dutta:2019gen, Akers:2021pvd} and the balanced partial entanglement \cite{Wen:2021qgx}.

On a separate note in the past a class of $(1+1)$-dimensional Galilean conformal field theories ($GCFT_{1+1}$s) was described in \cite{Bagchi:2009ca, Bagchi:2009my, Bagchi:2009pe} utilizing an \.In\"on\"u-Wigner contraction of the symmetry algebra for relativistic $CFT_{1+1}$s. Interestingly the EE for bipartite states in these $GCFT_{1+1}$s could be also computed through an appropriate replica technique described in  \cite{Bagchi:2014iea}. In subsequent works, the authors in \cite{Basu:2015evh, Jiang:2017ecm, Hijano:2017eii, Godet:2019wje} established a holographic construction to obtain the EE in the context of flat space holography \cite{Bagchi:2010zz, Bagchi:2013qva}. 

As described earlier, the EE fails to be a viable entanglement measure for bipartite mixed states. In this context, the issue of characterizing mixed state entanglement for bipartite states of these $GCFT_{1+1}$s assumes a critical significance. Addressing this issue, in \cite{Malvimat:2018izs} the authors had obtained the entanglement negativity for bipartite states through a replica technique. A holographic characterization of the entanglement negativity in the flat holographic framework was also recently described in \cite{Basu:2021axf}. The authors utilized the algebraic sums of the lengths of extremal curves for the dual bulk asymptotically flat geometries, homologous to certain combinations of the intervals relevant to the mixed state configuration in the $GCFT_{1+1}$. These constructions were motivated by earlier constructions in the literature for the usual $AdS_3/CFT_2$ scenario \cite{Chaturvedi:2016rcn, Jain:2017aqk, Malvimat:2018txq}. Furthermore the authors in \cite{Basu:2021awn} obtained the bulk EWCS for bipartite states in dual $GCFT_{1+1}$s through a novel geometric construction in the context of flat holography. Recently, the holographic duality between the bulk EWCS and the balanced partial entanglement \cite{Wen:2021qgx} was investigated and verified in \cite{Camargo:2022mme, Basu:2022nyl}. Also the authors in \cite{Basak:2022cjs} obtained the reflected entropy for bipartite states in $GCFT_{1+1}$s and compared their results with the EWCS to verify the duality between the EWCS and the reflected entropy described in \cite{Dutta:2019gen} (see also \cite{Setare:2022uid}).

The above developments naturally lead to the interesting issue of the computation for the OEE of bipartite states in $GCFT_{1+1}$ dual to bulk asymptotically flat geometries and explicitly verify the holographic duality of the bulk EWCS with the difference between the OEE and the EE in the context of  flat holography. In this article we address this significant issue and construct an appropriate replica technique to compute the OEE for bipartite states in $GCFT_{1+1}$s. To this end we first obtain the OEE for bipartite pure and mixed states in $CFT_{1+1}$s which are missing in the literature. Subsequently, using a replica technique, we obtain the OEE for bipartite states involving a single, two adjacent and two disjoint intervals in $GCFT_{1+1}$s at zero and finite temperatures and for finite sized systems. For the case of the two disjoint intervals we implement a geometric monodromy analysis \cite{Hijano:2018nhq} for the corresponding four point twist field correlator in the $GCFT_{1+1}$ to obtain the relevant dominant Galilean conformal block in the large central charge limit. Furthermore we compare our results to the bulk EWCS computed in \cite{Basu:2021awn} and explicitly verify the holographic duality with the difference between the OEE and the EE in flat holographic scenarios.

The rest of the article is organized as follows. In \cref{sec2} we describe the OEE and briefly review the corresponding replica technique for bipartite states in the context of  the usual relativistic $CFT_{1+1}$. We also utilize this replica technique to compute the OEE for certain bipartite states at zero and finite temperatures and in finite sized systems described by $CFT_{1+1}$s which were missing in the literature. Subsequently, in \cref{sec3}, after a brief review of the $(1+1)$-dimensional non-relativistic Galilean conformal field theories, we establish a replica technique to compute the OEE for various bipartite states in $GCFT_{1+1}$s and compare our results with the bulk EWCS. In \cref{sec4} we present a summary of our work and the conclusions. Finally in Appendix \ref{appendix_A} we provide a limiting analysis where we show that our result of the OEE for the bipartite mixed state of two disjoint interval is consistent with the appropriate non-relativistic limit of the corresponding $CFT_{1+1}$ result.

\section{OEE in conformal field theories}\label{sec2}

\subsection{Odd entanglement entropy}
We begin with a brief review of the odd entanglement entropy (OEE), introduced in \cite{Tamaoka:2018ned} as a correlation measures between the two subsystems of a bipartite mixed state. It can roughly be described as the von Neumann entropy of the partially transposed reduced density matrix of the given subsystem (cf. \cref{OEE-roughly-EE}). In this context, one starts with a tripartite pure state composed of the subsystems $A_1$, $A_2$ and $B$. Subsequently the subsystem $B$ is traced out to obtain the bipartite mixed state for the subsystem $A=A_1 \cup A_2$ with the reduced density matrix $\rho_{A_1A_2}$ defined on the Hilbert space $\mathcal{H} = \mathcal{H}_{A_1} \otimes \mathcal{H}_{A_2}$. The partial transposition of the reduced density matrix $\rho_{A_1A_2}$ with respect to the subsystem $A_2$ is defined as
\begin{equation}\label{partial-transpose}
\mel{e^{(1)}_ie^{(2)}_j}{\rho_{A_{1}A_{2}}^{T_{A_{2}}}}{e^{(1)}_ke^{(2)}_l} = 
\mel{e^{(1)}_ie^{(2)}_l}{\rho_{A_{1}A_{2}}}{e^{(1)}_ke^{(2)}_j},
\end{equation}
where ${|e^{(1)}_i\rangle}$ and $|e^{(2)}_j\rangle$ are the bases for the Hilbert spaces $\mathcal{H}_{A_1}$ and  $\mathcal{H}_{A_2}$ respectively. Further we define the  R\'enyi generalization of the OEE for the partially transposed density matrix as follows
\begin{equation}\label{odd-renyi}
	S_o^{(n_o)}\left(A_{1}:A_{2}\right)=\frac{1}{1-n_o}\log\left[\textrm{Tr}_{\mathcal{H}}(\rho_{A_{1}A_{2}}^{T_{A_{2}}})^{n_{0}}\right]	
\end{equation}
where $n_o$ is an odd integer\footnote{The partially transposed density matrix raised to an even power leads to another mixed state entanglement measure termed the entanglement negativity \cite{Vidal:2002zz}.}. The odd entanglement entropy $S_o$ for the given mixed state $\rho_{A_1 A_2}$ may finally be obtained through the analytic continuation of the odd integer $n_o \to 1$ in the above expression as follows\footnote{The author in \cite{Tamaoka:2018ned}, instead used the Tsallis entropy to obtain the OEE. However note that, in the replica limit $n_o \to 1$ both the R\'enyi generalization of the OEE in \cref{odd-renyi} and the Tsallis entropy considered in \cite{Tamaoka:2018ned} matches and gives the same expression for the OEE.} \cite{Tamaoka:2018ned}
\begin{equation}\label{oee-def}
S_o(A_{1}:A_{2})= \lim_{n_o \to 1}[S_{o}^{(n_o)}(A_{1}:A_{2})].
\end{equation}

In \cite{Mollabashi:2020ifv}, the authors numerically confirmed the following quantum information properties which ensures that the OEE is a well-defined bipartite mixed state measure:
\begin{itemize}
	\item $S_o(A:B) \ge 0$ (positive semi-definite)
	\item $S_o(A:B_1 B_2) \ge S_o(A:B_1)$ (monotonic)
	\item $S_o(A:B_1 B_2) \le S_o(A:B_1) +S_o(A:B_2)$ (polygamy relation) 
	\item $S_o(A_1 A_2:B_1 B_2) \ge S_o(A_1:B_1) +S_o(A_2:B_2)$ (breaking of strong super additivity).
\end{itemize}
However, a general analytic proof of these properties remain an open issue.

\subsubsection{OEE in holographic $CFT_{1+1}$}
\begin{figure}[H]
	\centering
	\includegraphics[scale=1]{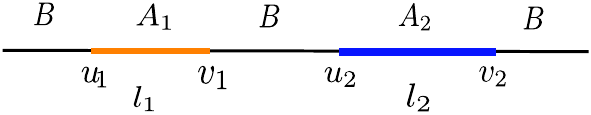}\\
	\caption{Two disjoint intervals $A_1$ and $A_2$.}
	\label{fig1}
\end{figure}

In this subsection, we review the replica technique utilized for the computation of the OEE in $(1+1)$-dimensional conformal field theories ($CFT_{1+1}$s) as described in \cite{Tamaoka:2018ned}. For the $CFT_{1+1}$s, the trace $\textrm{Tr}_{ \mathcal{H}} ( \rho_{A_{1} A_{2}}^{T_{A_{2}}} )^{n_{o}}$ in \cref{odd-renyi} may be expressed as a twist field correlator corresponding to the mixed state in question. We consider a generic tripartite pure state in a $CFT_{1+1}$ which is described by the intervals $A_{1}= [u_{1},v_{1}]$, $A_{2}= [u_{2},v_{2}] $ and  $B=(A_1 \cup A_2)^c$ as shown in \cref{fig1}. For the bipartite mixed state of $A_1 \cup A_2$ obtained by tracing out the degrees of freedom corresponding to the subsystem $B$, the trace $\textrm{Tr}_{ \mathcal{H}} (\rho_{A_{1} A_{2}}^{ T_{A_{2}}} )^{n_{o}}$ may be expressed as a four-point twist field correlator on the complex plane as follows \cite{Tamaoka:2018ned, Calabrese:2012nk}
\begin{equation}\label{four-point-correlator}
\textrm{Tr}_{\mathcal{H}}(\rho_{A_{1}A_{2}}^{T_{A_{2}}})^{n_{o}}=\left<\mathcal{T}_{n_{o}}(u_{1})\mathcal{\bar{T}}_{n_{o}}(v_{1})\mathcal{\bar{T}}_{n_{o}}(u_{2})\mathcal{T}_{n_{o}}(v_{2})\right>.
\end{equation}
Here $\mathcal{T}_{n_{o}}$ and $\mathcal{\bar{T}}_{n_{o}}$ are the twist and anti-twist field operators in $CFT_{1+1}$ respectively with the following weights,
\begin{equation}\label{conformal-weights}
h_{\mathcal{T}_{n_{o}}}=\bar{h}_{\mathcal{\bar{T}}_{n_{o}}}=\frac{c}{24} \left(n_{o} - \frac{1}{n_{o}}\right),
\end{equation} 
where $c$ is the central charge. The four point twist correlator in \cref{four-point-correlator} was utilized in \cite{Tamaoka:2018ned} to obtain the OEE for the bipartite mixed state of two disjoint intervals in a $CFT_{1+1}$ at zero temperature. In this article, we further utilize the above replica technique to obtain the OEE for the bipartite states of two disjoint intervals at a finite temperature and for a finite sized system. We also obtain the OEE for two adjacent and a single interval at zero and finite temperatures and for a finite sized system in $CFT_{1+1}$s.

In context of the $AdS_{3}/CFT_{2}$ correspondence, the authors in \cite{Tamaoka:2018ned} have also proposed a holographic duality for the difference of the OEE and the EE in terms of the bulk minimal entanglement wedge cross section (EWCS) corresponding to the bipartite state under consideration as follows
\begin{equation} \label{duality}
	S_o (A_1:A_2) - S (A_1 \cup A_2) = E_W (A_1:A_2)\, ,
\end{equation}
where $S(A_1 \cup A_2)$ denotes the EE and $E_W (A_1:A_2)$ denotes the minimal EWCS for the subsystem $A_1 \cup A_2$.

\subsection{OEE for two disjoint intervals}\label{sec:dis-CFT}
In the following subsections we first review the OEE for two disjoint intervals at zero temperature as computed in \cite{Tamaoka:2018ned}. Subsequently we compute the OEE for the bipartite mixed state configuration of two disjoint intervals in a finite sized system and at a finite temperature using the replica technique described earlier.

\subsubsection{Two disjoint intervals at zero temperature} \label{dis-zero-cft}
For this case, consider the mixed state configuration of two disjoint intervals $A_{1} \equiv [u_{1},v_{1}]$ and $A_{2} \equiv [u_{2},v_{2}]$ in the vacuum state of the $CFT_{1+1}$. This configuration is described by the four point twist correlator given in \cref{four-point-correlator} which may be expanded in terms of the conformal blocks in the $t$-channel as \cite{Tamaoka:2018ned}
\begin{equation}\label{block-expansion}
	\frac{\left<\mathcal{T}_n(u_{1})\mathcal{\bar{T}}_n(v_{1})\mathcal{\bar{T}}_n(u_{2})\mathcal{T}_n(v_{2})\right>}{\left(|u_{1}-v_{2}| |v_1-u_2|\right)^{(-\frac{c}{6}[\frac{n^2-1}{n}])}}= \sum_p b_p \,  \mathcal{F}(c,h_{\mathcal{T}_{n}},h_p,1-x) \,
	\mathcal{\bar{F}}(c,\bar{h}_{{\mathcal{\bar{T}}}_{n}},\bar{h}_p,1-\bar{x}),
\end{equation}
where $x= \frac{ \left|u_1-v_1\right| \left|u_2-v_2\right|}{ \left|u_1-u_2\right| \left|v_1-v_2\right| }$ is the cross ratio, $b_p$ is the OPE coefficient\footnote{The contribution due to the OPE coefficient is negligible in the large central charge limit as it is independent of the position of the primary operators \cite{Tamaoka:2018ned}.} and $\mathcal{F}$ and $\mathcal{\bar{F}}$ are the Virasoro conformal blocks corresponding to the exchange of the primary operator with the dimension $h_p$. In the $t$-channel, the dominant contribution arises from the twist field operator $\mathcal{T}_{n_o}^{2}$ with the weight $h_{\mathcal{T}_{n_{o}}}^{(2)} =h_{\mathcal{T}_{n_{o}}}$. The conformal block for this case may be expressed as \cite{Tamaoka:2018ned, Fitzpatrick:2014vua}
\begin{equation}
	\log \mathcal{F} (c,h_{\mathcal{T}_{n_{o}}}, h_{\mathcal{T}_{n_{o}}}^{(2)}, 1-x) = -h_{\mathcal{T}^{(2)}_{n_{o}}} \log \left[ \frac{1+ \sqrt{x}}{1- \sqrt{x}} \right].
\end{equation}
Using the above expression and \cref{block-expansion,oee-def}, one may obtain the OEE for the mixed state of two disjoint intervals in the $CFT_{1+1}$ as follows \cite{Tamaoka:2018ned}
\begin{equation}\label{Dis.}
	S_o(A_{1}:A_{2})= S(A_{1}\cup A_{2})+\frac{c}{6}\log \left[\frac{1+\sqrt{x}}{1-\sqrt{x}}\right]+\ldots ,
\end{equation}
where ellipsis denote subsequent terms which are sub-leading in the large central charge limit. The first term in the above expression denotes the EE for the subsystem $A_1 \cup A_2$ given as
\begin{equation}
	S(A_{1}\cup A_{2})=\frac{c}{3}\log\left(\frac{|u_1-v_2|}{a}\right)+\frac{c}{3}\log\left(\frac{|v_1-u_2|}{a}\right) + \ldots \, ,
\end{equation} 
where $a$ is a UV cut off of the $CFT_{1+1}$. It is straightforward to check that in the large central charge limit, \cref{Dis.} satisfies the duality in \cref{duality} as the second term matches exactly with the corresponding EWCS obtained in \cite{Takayanagi:2017knl}.

\subsubsection{Two disjoint intervals in a finite size system} \label{sec:disj-L-cft}
For this case, we consider the mixed state configuration of two disjoint intervals $A_1$ and $A_2$ of lengths $l_1$ and $l_2$ respectively, in a $CFT_{1+1}$ described on a cylinder with circumference $L$. To obtain the OEE for the given mixed state under consideration, it is necessary to compute the corresponding four point twist correlator in \cref{four-point-correlator} on the cylinder. This may be done by utilizing the following conformal map which maps the complex plane to a cylinder \cite{Calabrese:2012nk, Calabrese:2009qy}
\begin{equation}\label{Trans-size}
	z \to \omega= \frac{i L}{2\pi} \log{z},
\end{equation}
where $z$ describe the coordinates on the complex plane and $w$ describe the coordinates on the cylinder. Under this conformal map, the cross-ratio modifies as
\begin{equation}\label{mod.}
	\tilde{x}= \frac{\sin\left(\frac{\pi l_{1}}{L}\right)\sin\left(\frac{\pi l_2}{L}\right)}{\sin\left(\frac{\pi (l_{1}+l_{s})}{L}\right)\sin\left(\frac{\pi (l_2+l_s)}{L}\right)},
\end{equation}
where $l_s$ represents the length of the region sandwiched between the two intervals $A_1$ and $A_2$. Now, utilizing eqs. \eqref{Dis.}, \eqref{Trans-size} and \eqref{mod.}, we may obtain the OEE for the mixed state of two disjoint intervals in question as  
\begin{equation}\label{Result}
	S_o(A_{1}:A_{2})= \frac{c}{3}\log\left(\frac{L}{\pi a} \sin\frac{\pi(l_1+l_2+l_s)}{L}\right)+\frac{c}{3}\log\left(\frac{L}{\pi a}\sin\frac{\pi l_s}{L}\right)+\frac{c}{6}\log \left[\frac{1+\sqrt{\tilde{x}}}{1-\sqrt{\tilde{x}}}\right]+ \ldots ,
\end{equation}
where $a$ is a UV cut-off. Note that the first two terms in the above expression denote the EE $S (A_1 \cup A_2)$ for the subsystem $A_1 \cup A_2$. The corresponding bulk EWCS for this bipartite state may be obtained easily by utilizing \cref{Trans-size}. On computing the bulk EWCS we observe that it matches exactly with the last term in the above expression for the OEE. This provides substantiation to our computations as \cref{Result} is consistent with the holographic duality described in \cref{duality}.

\subsubsection{Two disjoint intervals at a finite temperature} \label{sec:disj-T-cft}
We now turn our attention to the mixed state configuration of two disjoint intervals at a finite temperature. To this end, we consider two disjoint intervals $A_1$ and $A_2$ of lengths $l_1$ and $l_2$ respectively, in a $CFT_{1+1}$ at a finite temperature defined on a thermal cylinder with circumference given by the inverse temperature $\beta$. To obtain the OEE for this case, it is required to obtain the four point twist correlator in \cref{four-point-correlator} on the thermal cylinder. We may employ the following conformal map to transform the complex plane to the thermal cylinder \cite{Calabrese:2009qy, Calabrese:2012nk}
\begin{equation}\label{Trans-temp}
	z \to \omega= \frac{\beta}{2\pi}
	\log{z},
\end{equation}
where $z$ denote the coordinates on the complex plane and $w$ denotes the coordinates on the thermal cylinder. The $CFT$ cross-ratio modifies under this transformation as
\begin{equation}\label{Mod.}
	\hat{x}= \frac{\sinh\left(\frac{\pi l_{1}}{\beta}\right)\sinh\left(\frac{\pi l_2}{\beta}\right)}{\sinh\left(\frac{\pi (l_{1}+l_{s})}{\beta}\right)\sinh\left(\frac{\pi (l_2+l_s)}{\beta}\right)}\,,
\end{equation} 
where, similar to the previous case, $l_s$ is the length of the region sandwiched between the disjoint intervals $A_1$ and $A_2$. Now using eqs. \eqref{Trans-temp}, \eqref{Mod.} and \eqref{Dis.} we may obtain the OEE for the mixed state of two disjoint intervals at a finite temperature as
\begin{equation}\label{result}
	S_o(A_{1}:A_{2})= \frac{c}{3}\log\left(\frac{\beta}{\pi a} \sinh\frac{\pi(l_1+l_2+l_s)}{\beta}\right)+\frac{c}{3}\log\left(\frac{\beta}{\pi a}\sinh\frac{\pi l_s}{\beta}\right)+\frac{c}{6}\log \left[\frac{1+\sqrt{\hat{x}}}{1-\sqrt{\hat{x}}}\right]+ \ldots.
\end{equation}
Again note that the first two terms correspond to the EE for the mixed state $A_1 \cup A_2$ and the last term matches exactly with the corresponding bulk EWCS obtained in \cite{Takayanagi:2017knl} which is consistent with the holographic duality \eqref{duality}.

\subsection{OEE for adjacent intervals}
Having computed the OEE for the various bipartite state configurations of two disjoint intervals, we now turn our attention to the OEE for the mixed states described by two adjacent intervals. 

\subsubsection{Adjacent intervals at zero temperature}
For the zero temperature case we consider two adjacent intervals $A_1 \equiv [-l_1,0]$ and $A_2 \equiv [0,l_2]$ in the $CFT_{1+1}$. This configuration may be obtained by taking the limit $v_1 \to u_2$ in \cref{four-point-correlator}. The trace $\textrm{Tr}(\rho_{A_{1}A_{2}}^{T_{A_{2}}})^{n_{o}}$ may then be obtained by the following three point twist correlator on the complex plane
\begin{equation}\label{three-point-correlator}
\textrm{Tr}(\rho_{A_{1}A_{2}}^{T_{A_{2}}})^{n_{o}}=\left<\mathcal{T}_{n_{o}}(-l_{1})\mathcal{\bar{T}}^2_{n_{o}}(0)\mathcal{T}_{n_{o}}(l_{2})\right>.
\end{equation}
Using the usual form of a three point correlator in a $CFT_{1+1}$, we may obtain the OEE for the mixed state in question by using \cref{oee-def} as follows
\begin{equation}\label{adj-cft}
S_o(A_{1}:A_{2})= \frac{c}{6} \log\left(\frac{l_1 l_2}{a^2}\right)+\frac{c}{6} \log\left(\frac{l_1+l_2}{a}\right)+\ldots,
\end{equation}
where $a$ is again a UV cut off for the $CFT_{1+1}$. Note that the above expression matches exactly with the corresponding OEE computed in \cite{Mollabashi:2020ifv} in the context of $1$-dimensional harmonic spin chain and in \cite{Dong:2021clv} for the gravitational path integral computation based on fixed area states\footnote{Note that in \cite{Dong:2021clv}, the OEE has been termed as the ``partially transposed entropy" due to the loose interpretation of the OEE being the analogue of the EE for the partially transposed density matrix.}. We may also rewrite the above expression as follows
\begin{equation}\label{adj-cft result}
S_o{(A_1:A_2)}-S(A_1 \cup A_2)= \frac{c}{6} \log(\frac{l_1 l_2}{a(l_{1}+l_{2})})+\ldots,
\end{equation}
where $S(A_1 \cup A_2)$ denotes the EE for the corresponding mixed state $A_1 \cup A_2$ given as 
\begin{equation}
S(A_1 \cup A_2)= \frac{c}{3} \log(\frac{l_{1}+l_{2}}{a})+\ldots.
\end{equation}
We observe here that the right-hand-side of \cref{adj-cft result} matches with the corresponding EWCS \cite{Kudler-Flam:2018qjo} apart from an additive constant which is contained in the OPE coefficient of the corresponding three point twist correlator in \cref{three-point-correlator}. We would also like to note here that the adjacent intervals configuration under consideration can also be obtained through an appropriate adjacent limit $v_1 \to u_2$ of the disjoint intervals configuration discussed in subsection \ref{dis-zero-cft}, and our results in \cref{Dis.,adj-cft} are consistent with this limiting behaviour.

\subsubsection{Adjacent intervals in a finite size system}
We now proceed to the mixed state of two adjacent intervals in a finite size system. For this case we consider the bipartite configuration involving two adjacent intervals $A_1$ and $A_2$ of lengths $l_1$ and $l_2$ respectively, in a $CFT_{1+1}$ defined on a cylinder with circumference $L$. We again employ the conformal transformation in \cref{Trans-size}, which maps the complex plane to the required cylinder of circumference $L$. We may now obtain the OEE for the mixed state configuration in question by using \cref{Trans-size,adj-cft} as follows
\begin{equation}\label{adjacent-L-cft}
S_o(A_{1}:A_{2})= \frac{c}{6} \log \left[\left(\frac{L}{\pi a}\right)^3 \sin\left(\frac{\pi l_{1}}{L}\right)\sin\left(\frac{\pi l_2}{L}\right) \sin\left(\frac{\pi (l_1+l_2)}{L}\right) \right]+\ldots,
\end{equation}
where $a$ is a UV cut-off. Again, the above result can also be obtained through the appropriate adjacent limit $v_1 \to u_2$ in the disjoint intervals result given in \cref{Result}. This serves as a yet another consistency check for our computations.

We may rewrite the above expression in the following way
\begin{equation}\label{adj-L-cft}
S_o{(A_1:A_2)}-S(A_1 \cup A_2)=\frac{c}{6} \log\left[\frac{L}{\pi a} \frac{\sin\left(\frac{\pi l_{1}}{L}\right)\sin\left(\frac{\pi l_{2}}{L}\right)}{\sin\left(\frac{\pi (l_{1}+l_{2})}{L}\right)}\right]+\ldots,
\end{equation}
where $S(A_1 \cup A_2)$ is the EE for the corresponding mixed state $A_1 \cup A_2$ given as
\begin{equation}
S(A_1 \cup A_2)=\frac{c}{3} \log\left[\frac{L}{\pi a}\sin\left(\frac{\pi (l_{1}+l_{2})}{L}\right)\right]+\ldots.
\end{equation}
Similar to the subsection \ref{sec:disj-L-cft}, we obtain the corresponding bulk EWCS for this case by utilizing \cref{Trans-size} and find that it matches with the right-hand-side of \cref{adj-L-cft}, modulo a constant contained in the undetermined OPE coefficient of the corresponding three point twist correlator.

\subsubsection{Adjacent intervals at a finite temperature}
For this case, we consider the mixed state configuration of two adjacent intervals $A_1$ and $A_2$ of lengths $l_1$ and $l_2$ respectively, in a $CFT_{1+1}$ at a finite temperature $T={1/\beta}$ defined on a thermal cylinder with circumference $\beta$. Similar to the finite size case in the previous subsection, we may compute the corresponding three point twist correlator given in \cref{three-point-correlator} on the thermal cylinder using the conformal map in \cref{Trans-temp}. Finally we may obtain the OEE for the two adjacent intervals at a finite temperature using \cref{oee-def} to be
\begin{equation}\label{adjacent-T-cft}
S_o(A_{1}:A_{2})= \frac{c}{6} \log \left[\left(\frac{\beta}{\pi a}\right)^3 \sinh\left(\frac{\pi l_{1}}{\beta}\right)\sinh\left(\frac{\pi l_2}{\beta}\right) \sinh\left(\frac{\pi (l_1+l_2)}{\beta}\right)\right]+\ldots,
\end{equation}
where $a$ is a UV cut off. We again note that the present bipartite configuration under consideration may also be obtained through an appropriate adjacent limit of the disjoint intervals configuration in subsection \ref{sec:disj-T-cft}, and our result in the above expression conforms to this limiting behaviour.

We may rewrite \cref{adjacent-T-cft} in the following way
\begin{equation}
S_o{(A_1:A_2)}-S(A_1 \cup A_2)=\frac{c}{6} \log\left[\frac{\beta}{\pi a} \frac{\sinh\left(\frac{\pi l_{1}}{\beta}\right)\sinh\left(\frac{\pi l_{2}}{\beta}\right)}{\sinh\left(\frac{\pi (l_{1}+l_{2})}{\beta}\right)}\right]+\ldots,
\end{equation}
where $S(A_1 \cup A_2)$ is the EE for the bipartite state $A_1 \cup A_2$ given as
\begin{equation}
	S(A_1 \cup A_2)=\frac{c}{3} \log\left[\frac{\beta}{\pi a}\sinh\left(\frac{\pi (l_{1}+l_{2})}{\beta}\right)\right]+\ldots.
\end{equation}
Similar to the previous cases, the expression for the corresponding EWCS is not present in the literature. However we expect that the above equation is consistent with the duality \eqref{duality} and the right-hand-side denotes the EWCS apart from the an additive constant arising from the OPE coefficient of the corresponding three point twist correlator. We again leave the explicit computation of the EWCS and the verification of our claim to future prospects.

\subsection{OEE for a single interval}
Having discussed the bipartite configurations involving two disjoint and adjacent intervals, we finally turn our attention to the OEE for bipartite pure and mixed states involving a single interval in $CFT_{1+1}$s.

\subsubsection{Single interval at zero temperature}
In this subsection, we consider the pure state of a single interval $A_1 \equiv [u_1,v_1]$ of length $l= |u_1-v_1|$ in the $CFT_{1+1}$, which may be obtained through the limit $u_2 \to v_1$ and $v_2 \to u_1$ in \cref{four-point-correlator}. For this case $A_1 \cup A_2$ describes the full system with $B$ as a null set. In this limit the four point twist correlator in \cref{four-point-correlator} reduces to the following two point twist correlator:
\begin{equation}
	\textrm{Tr}_{\mathcal{H}}(\rho_{A_{1}A_{2}}^{T_{A_{2}}})^{n_{o}}=\left<\mathcal{T}_{n_{o}}^{2}(u_{1})\mathcal{\bar{T}}_{n_{o}}^{2}(v_{1})\right>.
\end{equation}
Here the twist field operator $\mathcal{T}^{2}_{n_o}$ connects the $n_o$-th sheet with the $(n_o+2)$-th sheet and have dimensions $h_{\mathcal{T}_{n_{o}}}^{(2)} =h_{\mathcal{T}_{n_{o}}}$. As described in \cite{Calabrese:2012nk}, the above two point twist correlator is different for the even and the odd exponents of the trace $\textrm{Tr}_{\mathcal{H}} (\rho_{A_{1}A_{2}}^{T_{A_{2}}})^{n}$. For the present case where we have $n = n_o$ odd, the twist field operator $\mathcal{T}^{2}_{n_o}$ in the above two point twist correlator simply results in the reorganization of the $n_o$ replica sheets but does not change the structure of the $n_o$-sheeted Riemann manifold\footnote{Interestingly for even integers $n=n_e$, the $n_e$-sheeted Riemannian manifold decouples into two independent $(n_e/2)$-sheeted Riemann surfaces.} and hence we have the following \cite{Calabrese:2012nk}
\begin{equation} \label{single-two-point}
	\textrm{Tr}_{\mathcal{H}}(\rho_{A_{1}A_{2}}^{T_{A_{2}}})^{n_{o}}=\left<\mathcal{T}_{n_{o}}(u_{1})\mathcal{\bar{T}}_{n_{o}}(v_{1})\right>=\textrm{Tr}_{\mathcal{H}}(\rho_{A_{1}})^{n_{o}}.
\end{equation}
The OEE for the single interval in question may now be obtained using \cref{odd-renyi,oee-def} as
\begin{equation}\label{Single Int.result}
	S_o(A_{1}:A_{2})= \frac{c}{3} \log\left(\frac{|u_1-v_1|}{a}\right)+ \text{const.} ,
\end{equation}
where $a$ is a UV cut-off and the constant is due to the normalization of the two point twist correlator. As can be clearly seen, for the pure state configuration of the single interval in question, the OEE matches exactly with the EWCS \cite{Takayanagi:2017knl} which is identically equal to the the EE \cite{Calabrese:2004eu, Calabrese:2009qy}. This is in accordance with the expectation \cite{Tamaoka:2018ned} that for a pure state the OEE should reduce to the EE of the interval $A$ and the duality in \cref{duality} as the entropy $S(A_1 \cup A_2)$ vanishes for the state $A_1 \cup A_2$ describing the complete system.

\subsubsection{Single interval in a finite size system}
In this subsection we now focus on the configuration of a single interval in a finite sized system. To this end, we consider an interval $A_1$ of length $l$ with $A_2$ describing the rest of the system in a finite sized $CFT_{1+1}$ of length $L$ with periodic boundary condition. In this instance, the two point twist correlator in \cref{single-two-point} is required to be computed on a cylinder of circumference $L$. This is done by utilizing the conformal map given in \cref{Trans-size}. The OEE may then be obtained for the single interval in the finite sized system through \cref{single-two-point,odd-renyi,oee-def} as 
\begin{equation}\label{single-L-cft}
	S_o(A_{1}:A_{2})= \frac{c}{3} \log \left(\frac{L}{\pi a} \sin \frac{\pi l}{L}\right)+ \ldots .
\end{equation}
Similar to the previous case, we observe that the OEE for the single interval in question matches exactly with the corresponding EE for the interval $A_1$ \cite{Calabrese:2004eu, Calabrese:2009qy}. This is again in accordance with the duality \eqref{duality} as for this pure state configuration, $S(A_1 \cup A_2)=0$ and the EWCS reduces to the EE for $A_1$ \cite{Takayanagi:2017knl}.

\subsubsection{Single interval at a finite temperature}\label{sec:single-T-CFT}

For this case, we consider a single interval $A \equiv [-l,0]$ in a $CFT_{1+1}$ at a finite temperature $T$ defined on a thermal cylinder with circumference $\beta=1/T$. As described in \cite{Calabrese:2014yza} in the context of the entanglement negativity and in \cite{Basu:2022nds} in the context of the reflected entropy, it is necessary to consider two large but finite auxiliary intervals $B_1 \equiv [-L,-l]$, $B_2 \equiv [0,L]$ placed on either side of the interval $A$ in question. The OEE for the given single interval may then be obtained by utilizing the following four point twist correlator,
\begin{equation}\label{single-finite-T}
	S_o(A:B)=\lim_{L\to \infty} \lim_{n_o\to1}\frac{1}{1-n_o}\log \left[\left<\mathcal{T}_{n_o}(-L)\overline{\mathcal{T}}^2_{n_o}(-l)\mathcal{T}^2_{n_o}(0)\overline{\mathcal{T}}_{n_o}(L)\right>_\beta\right],
\end{equation}
where the subscript $\beta$ denotes that the twist correlator is being evaluated on the thermal cylinder with circumference $\beta$. Note that the bipartite limit $B \equiv B_1 \cup B_2 \to A^c$ ($L \to \infty$) has to be applied after the replica limit $n_o \to 1$ in the above equation as described in \cite{Calabrese:2014yza}. On the complex plane, the four point twist correlator described in eq. \eqref{single-finite-T} can be expressed as
\cite{Calabrese:2014yza}  
\begin{equation}\label{Cardy}
	\left<\mathcal{T}_{n_o}(z_1)\overline{\mathcal{T}}^2_{n_o}(z_2)\mathcal{T}^2_{n_o}(z_3)\overline{\mathcal{T}}_{n_o}(z_4)\right>_\mathbb{C}=\frac{k_{n_{o}}}{z_{14}^{2h_{\mathcal{T}_{n_{o}}}}z_{23}^{2h_{\mathcal{T}_{n_{o}}}^{(2)}}}\frac{\mathcal{F}_{n_o}(x)}{x^{h_{\mathcal{T}_{n_{o}}}^{(2)}}},
\end{equation}
where $k_{n_{o}}$ is a constant, $x=\frac{z_{12}z_{34}}{z_{13}z_{24}}$ is the cross ratio and $\mathcal{F}_{n_o}(x)$ is an arbitrary non universal function of the cross ratio. This non universal function in the limits $x \to 1$ and $x \to 0$ may be given as \cite {Calabrese:2014yza}
\begin{equation}
	\mathcal{F}_{n_o}(1)= 1, \hspace{2cm} \mathcal{F}_{n_o}(0)= C_{n_{o}},
\end{equation}
where $C_{n_{o}}$ is a non universal constant which depends on the full operator content of the field theory. The four point twist correlator in \cref{single-finite-T} on the thermal cylinder may be obtained by utilizing the transformation in \cref{Trans-temp}. The OEE for the mixed state configuration of the single interval in question may then be obtained using \cref{odd-renyi,oee-def} as
\begin{equation}
	S_o(A:B)= \lim_{L\to \infty}\left[\frac{c}{3}\log\left(\frac{\beta}{\pi a}\sinh\frac{2\pi L}{\beta}\right)\right]
	+\frac{c}{3}\log\left(\frac{\beta}{\pi a}\sinh\frac{\pi l}{\beta}\right)-\frac{\pi c l}{3\beta}+f\left(e^{-\frac{2\pi l}{\beta}}\right) + \ldots,
\end{equation}
where $a$ is a UV cut-off for the $CFT_{1+1}$ and the non-universal function $f(x) = \lim_{n_o \to 1} \ln \left [ \mathcal{F}_{n_o}(x) \right]$. Note that in the above expression first divergent term denotes the entanglement entropy $S(A\cup B)$ of total thermal system $A \cup B$ where the bipartite limit $B_1 \cup B_2 \to A^c$ has been taken. The finite part of the OEE may be extracted by subtracting this divergent EE as follows
\begin{equation}\label{single-T-cft}
	S_o(A:B)-S{(A \cup B)}= 	\frac{c}{3}\log\left(\frac{\beta}{\pi a}\sinh\frac{\pi l}{\beta}\right)-\frac{\pi c l}{3\beta}+f\left(e^{-\frac{2\pi l}{\beta}}\right)+ \ldots.
\end{equation}
It is instructive to express the above equation as
\begin{equation}
	S_o(A:B)-S(A \cup A^c) = S(A)-S^\text{th}(A) + f\left(e^{-\frac{2\pi l} {\beta}}\right)+ \ldots,
\end{equation} 
where $S(A)$ in the EE of the interval $A$ and $S^\text{th}(A)$ denotes the thermal entropy. We note here that the OEE evaluated in \cref{single-T-cft} matches exactly with the corresponding EWCS in \cite{KumarBasak:2020eia} in the large central charge limit. It is also worth pointing out that our result in \cref{single-T-cft} matches with the corresponding expressions for the OEE obtained in certain limits of the inverse temperature $\beta$ in \cite{Tamaoka:2018ned}. These serve as consistency checks for our computations.

\section{OEE in Galilean conformal field theories}\label{sec3}
Having discussed the computation of the OEE for various bipartite states in relativistic $CFT_{1+1}$s, in this section we now proceed to the analysis of the OEE in $(1+1)$-dimensional Galilean conformal field theories ($GCFT_{1+1}$s). We start with a short review of the $GCFT_{1+1}$s and subsequently compute the OEE for bipartite states involving two disjoint, two adjacent and a single interval in $GCFT_{1+1}$s.

\subsection{Review of  $GCFT_{1+1}$} \label{sec:review}
In this subsection, we briefly review certain essential features of $GCFT_{1+1}$ as described in \cite{Bagchi:2009my, Bagchi:2009pe, Bagchi:2009ca}. The Galilean conformal algebra ($GCA_{1+1}$) for $GCFT_{1+1}$s may be obtained through a parametric \.In\"on\"u-Wigner contraction of the usual Virasoro algebra which involves the rescaling of the space and the time coordinates as follows
\begin{equation}\label{Inonu-Wigner}
t\to t,\qquad   x_i\to \epsilon x_i,
\end{equation}
with $\epsilon\to 0$ which implies a vanishing velocity limit $v_i \sim \epsilon$. The action of a generic Galilean conformal transformation on the coordinates is equivalent to diffeomorphisms and $t$-dependent shifts respectively as follows 
\begin{equation}\label{finitetransf}
t\rightarrow f(t)\,,\qquad x\rightarrow f^{\prime} (t)\, x + g(t)\,.
\end{equation}
The generators of the $GCA_{1+1}$ in the plane representation are given as \cite{Bagchi:2009my}
\begin{equation}\label{GCGT Gen}
L_n=t^{n+1} \, \partial_{\,t}+(n+1) \, t^n \, x\, \partial_{\,x}, \qquad M_n=t^{n+1} \, \partial_x.
\end{equation}
The corresponding Lie algebra for the generators are then expressed as follows
\begin{equation}
\begin{aligned}
    \left[L_n,L_m\right]&= (m-n)L_{n+m}+\frac{C_L}{12}(n^3-n)\delta_{n+m,0},
    \\ [L_n,M_n]&=(m-n)M_{n+m}+\frac{C_M}{12}(n^3-n)\delta_{n+m,0},\\
    [M_n,M_m]&=0,
\end{aligned}
\end{equation}
where we have different central extensions for each sector involving the central charges $C_L$ and $C_M$ for the $GCFT_{1+1}$. The reduction of the Lorentz invariance to a Galilean invariance for the $GCFT_{1+1}$ results in two separate components for the energy-momentum tensor \cite{Hijano:2018nhq} and are given as
\begin{equation}\label{stressTensors}
\mathcal{M}\equiv T_{tx}=\sum_n M_n\,t^{-n-2}\quad,\quad \mathcal{L}\equiv T_{tt}=\sum_n\left[L_n+(n+2)\frac{x}{t}M_n\right]\,t^{-n-2} \, ,
\end{equation}
where the GCA generators $M_n$ and $L_n$ are defined in \cref{GCGT Gen}. The non-relativistic Ward identities for these two components $\mathcal{M}$ and $\mathcal{L}$ are given as \cite{Hijano:2018nhq}
\begin{equation}\label{WardIdentity}
\begin{aligned}
&\left<\mathcal{M}(x,t)V_1(x_1,t_1)\dots V_n(x_n,t_n)\right>=\sum_{i=1}^n \left[\frac{h_{M,i}}{(t-t_i)^2}+\frac{1}{t-t_i}\partial_{x_i}\right]\left< V_1(x_1,t_1)\dots V_n(x_n,t_n)\right>\,,\\
&\left<\mathcal{L}(x,t)V_1(x_1,t_1)\dots V_n(x_n,t_n)\right>=\sum_{i=1}^n \bigg[\frac{h_{L,i}}{(t-t_i)^2}-\frac{1}{t-t_i}\partial_{t_i}+\frac{2h_{M,i}(x-x_i)}{(t-t_i)^3}\\
&\qquad\qquad\qquad\qquad\qquad\qquad\qquad\qquad\quad+\frac{x-x_i}{(t-t_i)^2}\partial_{x_i}\bigg]\left< V_1(x_1,t_1)\dots V_n(x_n,t_n)\right>\,,
\end{aligned}
\end{equation}
where $V_i$s are  $GCFT_{1+1}$ primaries and $(h_{L,i},\, h_{M,i})$ are their corresponding weights. 

The usual form of a two point correlator of primary fields $V_i(x_i,t_i)$ may be obtained by utilizing the Galilean conformal symmetry as follows \cite{Bagchi:2009ca}
\begin{equation}\label{gcft two point}
   \big <V_1(x_1,t_1)V_2(x_2,t_2)\big >=C^{(2)}\delta_{h_{L,1}h_{L,2}}
   \delta_{h_{M,1}h_{M,2}}t_{12}^{-2h_{L,1}}\exp\left(-2h_{M,1}\frac{x_{12}}{t_{12}}\right),
\end{equation}
where  $x_{ij}=x_i-x_j$, $t_{ij}=t_i-t_j$ and $C^{(2)}$ is the normalization constant. In a similar manner the three point correlator of the primary fields could be expressed as \cite{Bagchi:2009ca}
\begin{equation}\label{gcft-3-point}
\begin{aligned}
\left<V_1(x_1,t_1)V_2(x_2,t_2)V_3(x_3,t_3)\right>=&C^{(3)}t_{12}^{-(h_{L,1}+h_{L,2}-h_{L,3})}\,
t_{23}^{-(h_{L,2}+h_{L,3}-h_{L,1})}\, t_{13}^{-(h_{L,1}+h_{L,3}-h_{L,2})}\times
\\
&\exp\Big[-(h_{M,1}+h_{M,2}-h_{M,3})\frac{x_{12}}{t_{12}}    -(h_{M,2}+h_{M,3}-h_{M,1}) \frac{x_{23}}{t_{23}}\\
&\qquad \qquad \qquad \qquad \qquad \quad \quad \qquad ~ -(h_{M,1}+h_{M,3}-h_{M,2}) \frac{x_{13}}{t_{13}}\Big],
\end{aligned}
\end{equation}
where $C^{(3)}$ is the OPE coefficient. Utilizing the Galilean symmetry, one may also express the four point correlator for primary fields $V_i(x_i,t_i)$ as \cite{Bagchi:2009ca, Malvimat:2018izs}
\begin{equation}\label{GCFT four point}
\left<\prod_{i=1}^4 V_i(x_i,t_i)\right>=\prod_{1\le i <j \le 4} t_{ij}^{\frac{1}{3}\sum_{k=1}^{4} h_{L,k}-h_{L,i}-h_{L,j}} \text{e}^{-\frac{x_{ij}}{t_{ij}}\left (\frac{1}{3} \sum_{k=1}^{4} h_{M,k}-h_{M,i}-h_{M,j}\right) } \mathcal{G}\left(T,\frac{X}{T}\right),       
\end{equation}
where  $\mathcal{G}(T,\frac{X}{T})$ is a non universal function which depend on the specific operator content of the $GCFT_{1+1}$. The non-relativistic cross-ratios $X$ and $\frac{X}{T}$ of the $GCFT_{1+1}$ are given as 
\begin{equation}\label{GCFT cross ratio}
T=\frac{t_{12}t_{34}}{t_{13}t_{24}}\,,\qquad \qquad \frac{X}{T}=\frac{x_{12}}{t_{12}}+\frac{x_{34}}{t_{34}}-\frac{x_{13}}{t_{13}}-\frac{x_{24}}{t_{24}}\,.
\end{equation}

In the following subsections, we will now obtain the OEE for various bipartite states in $GCFT_{1+1}$s involving two disjoint, two adjacent and a single interval.

\subsection{OEE for two disjoint intervals} \label{sec:disj-gcft}
\begin{figure}[H]
	\centering
	\includegraphics[scale=.7]{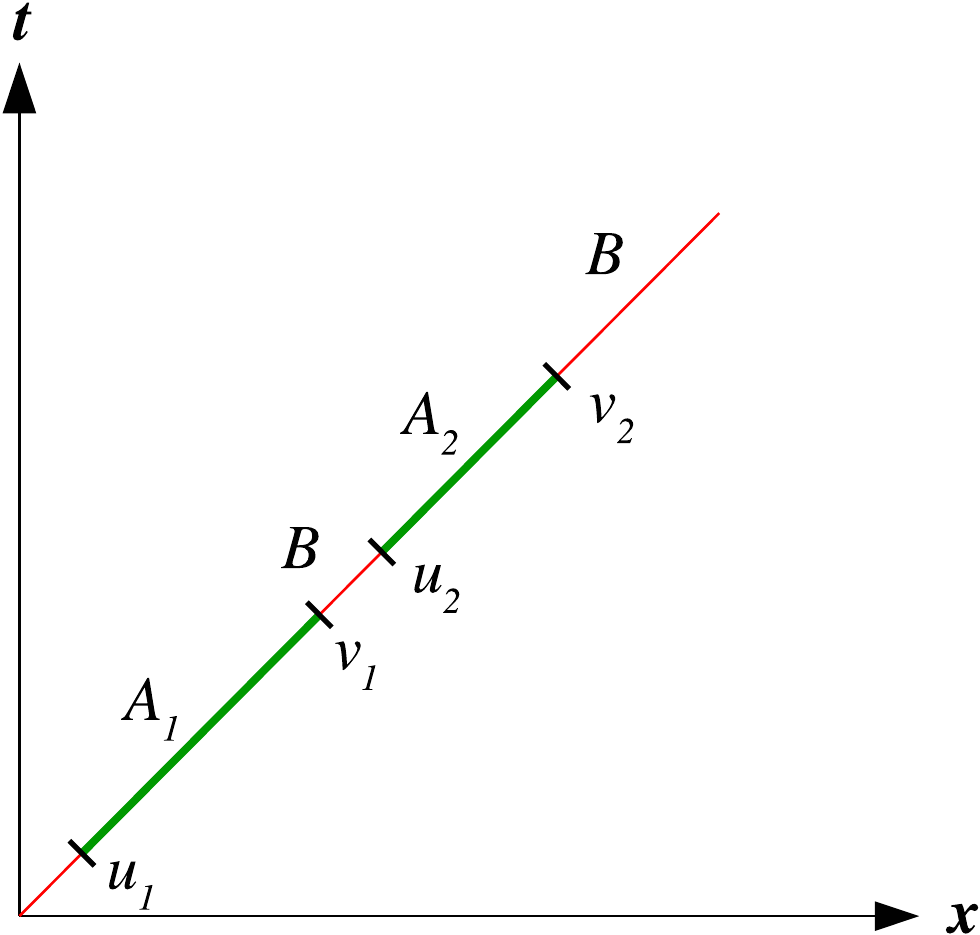}\\
	\caption{Boosted intervals $A_1$ and $A_2$ in a  $GCFT_{1+1}$ plane.}
	\label{fig2}
\end{figure}

As the $GCFT_{1+1}$ lacks the Lorentz invariance the OEE will be necessarily frame dependent and hence we need to consider Galilean boosted intervals henceforth. Similar to the relativistic case, the trace $\textrm{Tr}_{ \mathcal{H}} (\rho_{A_{1} A_{2}}^{T_{A_{2}}})^{n_{o}}$ required to obtain the OEE, can be computed through a replica technique and may be expressed as a twist correlator in the $GCFT_{1+1}$. Now let us consider the mixed state configuration of two disjoint boosted intervals $A_1 \equiv [u_1,v_1]$ and $A_2 \equiv [u_2,v_2]$ with $B$ describing the rest of the system as shown in \cref{fig2}. Here $u_1=(x_1,t_1),\, v_1=(x_2,t_2), u_2=(x_3,t_3),\, v_2=(x_4,t_4)$ are the end points of the intervals $A_1$ and $A_2$ respectively.
Similar to the case described in \cite{Malvimat:2018izs, Basu:2021axf} where the trace of the partially transposed density matrix is raised to an even exponent, the trace $\textrm{Tr}_{ \mathcal{H}} (\rho_{A_{1} A_{2}}^{T_{A_{2}}})^{n_{o}}$ with an odd exponent $n_o$ may be expressed as
\begin{equation}\label{trace}
	\textrm{Tr}_{\mathcal{H}}(\rho_{A_{1}A_{2}}^{T_{A_{2}}})^{n_{o}}=\left<\Phi_{n_{o}}(u_{1})\Phi_{-n_{o}}(v_1)\Phi_{-n_{o}}(u_2)\Phi_{n_{o}}(v_2)\right>,
\end{equation}
where $\Phi_{n_{o}}$ and $\Phi_{-n_{o}}$ are the twist and anti-twist operators in the $GCFT_{1+1}$, respectively with the corresponding weights given as
\begin{equation}\label{weights in GCFT}
	h_{L}^{(1)}=\frac{C_L}{24}\left(n_o-\frac{1}{n_o}\right)\, , \qquad
	h_{M}^{(1)}=\frac{C_M}{24}\left(n_o-\frac{1}{n_o}\right).
\end{equation}
The four point twist correlator of \cref{trace} is expected to exponentiate in the large central charge limit \cite{Basu:2021axf}, similar to the relativistic case discussed in subsection \ref{sec:dis-CFT}. The dominant contribution to the four point twist correlator may thus be extracted through the geometric monodromy analysis as discussed in \cite{Hijano:2017eii, Basu:2021axf}. To this end, in the large $C_L, C_M$ limit we may express the four point twist correlator in \cref{trace} in terms of the Galilean conformal block $\mathcal{F}_{\alpha}$ corresponding to the $t$-channel ($T \to 1 $, $X \to 0$) as follows 
\begin{equation}\label{4pointmonodromy}
	\begin{aligned}
		&\left<\Phi_{n_o}(x_1, t_1)\,\Phi_{-n_o}(x_2, t_2)\,\Phi_{-n_o}(x_3, t_3)\,\Phi_{n_o}(x_4, t_4)\right>\\
		&\qquad \qquad \qquad \qquad \qquad \quad = t_{14}^{-2h_{L}^{(1)}} \, t_{23}^{-2h_{L}^{(1)}} \, \exp\left[-2h_{M}^{(1)} \frac{x_{14}}{t_{14}}-2h_{M}^{(1)} \frac{x_{23}}{t_{23}}\right]\,\mathcal{F}_{\alpha} \left( T, \frac{X}{T} \right).
	\end{aligned}
\end{equation}
The dominant conformal block $\mathcal{F}_{\alpha}$ is an arbitrary function of the cross-ratios $X$ and $T$ and depends on the full operator content of the $GCFT_{1+1}$.

In the following subsections we obtain the expression of the block $\mathcal{F}_{\alpha}$ in the large central charge limit by utilizing a geometric monodromy analysis \cite{Hijano:2018nhq, Basu:2021axf} for each of the two components of the energy-momentum tensor $\mathcal{M}$ and $\mathcal{L}$ described in \cref{stressTensors}.

\subsubsection{Monodromy of $\mathcal{M}$} \label{sec:M-monodromy}
In this subsection, through the geometric monodromy analysis \cite{Hijano:2018nhq, Basu:2021axf} of the energy-momentum tensor component $\mathcal{M}$, we will obtain a partial expression for the Galilean conformal block $\mathcal{F}_{\alpha}$ in \cref{4pointmonodromy}. In this context, utilizing the Ward identities described in eq. \eqref{WardIdentity}, we obtain the expectation value of the $\mathcal{M}$ as
\begin{equation}
	\mathcal{M}(u_i;(x,t))= \sum_{i=1}^4\left[\frac{h_{M,i}}{(t-t_i)^2}+\frac{C_M}{6}\frac{c_i}{t-t_i}\right],
\end{equation}
where $u_i$ are the points in the $GCFT$ plane where the twist operators $\Phi_{n_o}$ are located and the auxiliary parameters $c_i$ are given by
\begin{equation}\label{auxC}
	c_i=\frac{6}{C_M} \, \partial_{x_i} \log \left<\Phi_{n_o} (u_1)\, \Phi_{-n_o} (u_2)\, \Phi_{-n_o} (u_3)\, \Phi_{n_o} (u_4)\right>\,.
\end{equation}
Note that for a four point correlator, the Galilean conformal symmetry is not sufficient to fix the structure of the correlator and hence some on the auxiliary parameters $c_i$ remain undetermined. By utilizing a Galilean transformation, we locate the twist operators at $t_1=0, \,t_3=1, \,t_4=\infty$ and leave $t_2=T$ free. We may express three of the auxiliary parameters in terms of the fourth by utilizing the facts that the expectation value of $\mathcal{M}$ scale as $\mathcal{M}(T;t) \sim \,t^{-4}$ as $t \to \infty$ and the conformal dimension $h_{M,i} \equiv h_M^{(1)}$ of the light operator $\Phi_{n_o}$ vanishes in the replica limit $n_o \to 1$. The expectation value of $\mathcal{M}$ may now be written in terms of the only unknown auxiliary parameter $c_2$ as follows \cite{Basu:2021axf}  
\begin{equation}
	\frac{6}{C_M}\mathcal{M}(T;t)=c_2\left[\frac{T-1}{t}+\frac{1}{t-T}-\frac{T}{t-1}\right]\,.
\end{equation}
Under a generic Galilean transformation given in eq. \eqref{finitetransf} the energy-momentum tensor $\mathcal{M}$ transforms as \cite{Hijano:2018nhq}
\begin{equation}
	\mathcal{M}^{\prime}(t^{\prime},x^{\prime})=(f^{\prime})^2\mathcal{M}(t,x)+\frac{C_M}{12}\,S(f,t)\,,
\end{equation}
where $S(f,t)$ is the Schwarzian derivative for the coordinate transformation $t \to f(t)$. For the ground state of the $GCFT_{1+1}$, the expectation value $\mathcal{M}(u_i;(x,t))$ vanishes on the $GCFT$ complex plane which constrains the Schwarzian derivative to have the following form
\begin{equation}\label{Schwarzian}
	\frac{1}{2}\,S(f,t)=c_2\left[\frac{T-1}{t}+\frac{1}{t-T}-\frac{T}{t-1}\right]\,.
\end{equation}
It is possible to express the above in the form of a differential equation as \cite{Basu:2021axf}
\begin{equation}\label{diff.eq}
	0= h^{\prime\prime}(t)+\frac{1}{2}S(f,t)\,h(t)=h^{\prime\prime}(t)+\frac{6}{c_M}\mathcal{M}(T,t)\,h(t)\,,
\end{equation}
where $f=h_1/h_2$ with $h_1$ and $h_2$ as the two solutions of the differential equation. By utilizing the monodromy of the solutions $h_1$ and $h_2$ by circling around the light operators at $t=1\,,T$ and using the monodromy condition for the three point twist correlator described in \cite{Basu:2021axf}, we may obtain the auxiliary parameter $c_2$ as
\begin{equation}
\quad c_2=\epsilon_{\alpha}\frac{1}{\sqrt{T}(T-1)}\,,
\end{equation}
where $\epsilon_{\alpha} = \frac{6}{C_M} h_{M,\alpha}$ is the rescaled weight of the corresponding conformal block $\mathcal{F}_\alpha$. We may now obtain the conformal block $\mathcal{F}_\alpha$ for the four-point function in eq. \eqref{4pointmonodromy} as
\begin{equation}\label{BLOCK}
	\begin{aligned}
		\mathcal{F}_{\alpha}&=\exp\left[\frac{C_M}{6}\int\,c_2\,dX\right]\\
		&=\exp\left[h_{M,\alpha}\left(\frac{X}{\sqrt{T}(T-1)}\right)\right]\tilde{\mathcal{F}}(T)\,.
	\end{aligned}
\end{equation}
Note that the complete form of the conformal block is still not known and we have an unknown function $\tilde{\mathcal{F}}(T)$ of the coordinate $T$ in the above expression. The form of this function will be determined through the geometric monodromy  analysis for the other component of the energy-momentum tensor $\mathcal{L}$ in the proceeding subsection.

\subsubsection{Monodromy of $\mathcal{L}$}
Similar to the Monodromy of $\mathcal{M}$ in the previous subsection, we may express the expectation value of $\mathcal{L}$ by utilizing the Galilean Ward identities in \cref{WardIdentity} as follows
\begin{equation}\label{Lexpect1}
	\frac{6}{C_M}\mathcal{L}(u_i;(x,t))=\sum_{i=1}^4\bigg[\frac{\delta_i}{(t-t_i)^2}-\frac{1}{t-t_i}d_{i}+\frac{2\epsilon_i(x-x_i)}{(t-t_i)^3}+\frac{x-x_i}{(t-t_i)^2}c_{i}\bigg]\,,
\end{equation}
where $\delta_i = \frac{6}{C_M} \, h_{L,i}$, $\epsilon_i = \frac{6}{C_M} \, h_{M,i}$, the auxiliary parameters $c_i$ are defined in eq. \eqref{auxC} and the auxiliary parameters $d_i$ are given as \cite{Hijano:2018nhq}
\begin{equation}\label{auxiliarydi}
	\begin{aligned}
		d_i=\frac{6}{C_M} \, \partial_{t_i} \log \left< \Phi_{n_o} (u_1)\, \Phi_{-n_o} (u_2)\, \Phi_{-n_o} (u_3)\, \Phi_{n_o} (u_4) \right>\,.
	\end{aligned}
\end{equation}
By utilizing a Galilean conformal map, we locate the twist operators at $t_1=0,\, t_2=T,\, t_3=1,\, t_4=\infty$ and $x_1=0,\, x_2=X,\, x_3=0$ and $x_4=0$. Again three of the four auxiliary parameters $d_i$ may be obtained in terms of the remaining one by utilizing the scaling $\mathcal{L}(T,t)\rightarrow t^{-4}$ with $t\rightarrow \infty$. We may now rewrite \cref{Lexpect1} in terms of the undetermined auxiliary parameter $d_2$ as follows
\begin{equation}
	\begin{aligned}
		\frac{6}{C_M}\mathcal{L}(u_i;(x,t))=&-\frac{c_2 X+d_2 (T-1)-2 \delta _L}{t}+\frac{c_2 X+d_2 T-2 \delta _L}{t-1}+\frac{c_1 x}{t^2}+\frac{c_2 (x-X)}{(t-T)^2}+\frac{c_3 x}{(t-1)^2}\\
		&-\frac{d_2}{t-T}+\frac{2 x \epsilon _L}{t^3}+\frac{\delta
			_L}{t^2}+\frac{\delta _L}{(t-1)^2}+\frac{\delta _L}{(t-T)^2}+\frac{2 \epsilon _L
			(x-X)}{(t-T)^3}+\frac{2 x \epsilon _L}{(t-1)^3}\,.\\
	\end{aligned}
\end{equation}
where $\delta_L=\frac{6}{C_M} h_{L}^{(1)}$ and $\epsilon_L=\frac{6}{C_M}h_{M}^{(1)}$ are the rescaled weights of the twist operator $\Phi_{n_o}$. We note here that the auxiliary parameters $c_i$ appearing in the above expression are as obtained in the preceding subsection \ref{sec:M-monodromy}.

As in \cite{Hijano:2017eii, Basu:2021axf}, now we consider the following combination of the expectation values of the two components of the energy-momentum tensor, 
\begin{equation}
	\mathcal{\tilde{L}} (u_i; (x,t))= \left[ \mathcal{L} (u_i; (x,t)) + X\, \mathcal{M}' (u_i; (x,t)) \right].
\end{equation}
By choosing an ansatz $g(t)=f'(t) Y(t)$ in the generic Galilean transformation \cref{finitetransf}, we get a differential equation of the following form,
\begin{equation}\label{Ltilda}
	\frac{6}{C_M}\mathcal{\tilde{L}}=-\frac{1}{2}Y'''-2 Y'\frac{6}{C_{M}}\mathcal{M}-Y\frac{6}{C_M}\mathcal{M'}\,.
\end{equation}
Similar to previous subsection, through the monodromy of the solutions of the above differential equation by circling around the light operators at $t=1\, , T$, we may determine the auxiliary parameter $d_2$ as 
\begin{equation}
	d_2= \frac{(1-3 T) X \epsilon _{\alpha }+2 (T-1) T \delta _{\alpha }}{2 (T-1)^2 T^{3/2}}\,,
\end{equation}
where $\epsilon _{\alpha }$, $\delta_{\alpha} = \frac{6}{C_M} \, h_{L,\alpha}$ are the rescaled weight of $\mathcal{F}_{\alpha}$. The complete form of the Galilean conformal block $\mathcal{F}_{\alpha}$ may now be obtained by using \cref{BLOCK,auxiliarydi} to be
\begin{equation}\label{Gconformalblock}
	\mathcal{F}_{\alpha} =\left(\frac{1-\sqrt{T}}{1+\sqrt{T}}\right)^{h_{L,\alpha}} \exp\left[h_{M,\alpha}\left(\frac{X}{\sqrt{T }(T-1)}\right)\right]\,.
\end{equation}

Now utilizing the above expression of the Galilean conformal block, we will obtain the OEE for the bipartite mixed states involving two disjoint intervals in $GCFT_{1+1}$s at zero and finite temperature and for a finite sized system.

\subsubsection{Two disjoint intervals at zero temperature}
In this subsection we compute the OEE for the bipartite mixed state of two disjoint intervals described by $A_1$ and $A_2$ in a $GCFT_{1+1}$ at zero temperature. To this end, we utilize the $t$-channel $T\to 1\,,X\to 0$ result in the large $C_M, C_L$ limit for the Galilean conformal blocks $\mathcal{F}_{\alpha}$ in \cref{Gconformalblock}. In particular, the dominant conformal block for the four point twist correlator in \cref{4pointmonodromy} in the $t$-channel is described by the primary twist field operator $\Phi^2_{n_o}$ with the weights $h_{L}^{(2)}=h_{L}^{(1)}$ and $h_{M}^{(2)}=h_{M}^{(1)}$ where $h_{L}^{(1)}, h_{M}^{(1)}$ are as given in \cref{weights in GCFT}. Utilizing the expression for the Galilean conformal block in \cref{Gconformalblock}, we may obtain the OEE for the two disjoint intervals under consideration as follows 
\begin{equation}\label{Dis. result(GCFT)}
\begin{aligned}
S_o(A_{1}:A_{2})  =&\frac{C_L}{6} \log\left(\frac{t_{14}t_{23}}{a^2}\right)+ \frac{C_L}{12} \log\left(\frac{1+\sqrt{T}}{1-\sqrt{T}}\right)\\
& + \frac{C_M}{6} \left(\frac{x_{14}}{t_{14}}+\frac{x_{23}}{t_{23}}\right)+\frac{C_M}{12} \frac{X}{\sqrt{T}} \left(\frac{1}{1-T}\right)
+\ldots,
\end{aligned}
\end{equation}
where $X$ and $T$ are the non relativistic cross ratios given in eq. \eqref{GCFT cross ratio}. The first and third term in the above expression denote the EE $S(A_1 \cup A_2)$ for the subsystem $A_1 \cup A_2$. The above expression may then be rearranged to obtain 
\begin{equation}\label{Dis-result}
S_o(A_{1}:A_{2})-S{(A_1 \cup A_2)}= \frac{C_L}{12} \log\left(\frac{1+\sqrt{T}}{1-\sqrt{T}}\right)+\frac{C_M}{12} \frac{X}{\sqrt{T}} \left(\frac{1}{1-T}\right)+\ldots,
\end{equation}
where the right-hand-side of the above expression matches exactly with the corresponding EWCS \cite{Basu:2021awn} in the context of flat holography. This verifies the duality described in \cref{duality} for flat holographic scenarios as well.

\subsubsection{Two disjoint intervals in a finite size system} \label{sec:dis-L-gcft}
Here we compute the OEE for the bipartite configuration of two disjoint intervals in a finite sized $GCFT_{1+1}$ defined on a cylinder of circumference $L$. In this context we consider the two disjoint intervals to be given as $A_1 \equiv [(\xi_1, \rho_1), (\xi_2, \rho_2)]$ and $A_2 \equiv [(\xi_3, \rho_3), (\xi_4, \rho_4)]$. Similar to the relativistic case in subsection \ref{sec:disj-L-cft}, it is required to calculate the four point twist correlator in \cref{4pointmonodromy} on the given cylinder. We utilize the following conformal map to transform from the $GCFT$ complex plane to the cylinder \cite{Malvimat:2018izs, Basu:2021axf}
\begin{equation}\label{Trans-size-gcft}
t_i=e^{\frac{2\pi i\xi_i}{L}}, \qquad x_i=\frac{2\pi i\rho_i}{L}e^{\frac{2\pi i\xi_i}{L}},
\end{equation}
where the coordinates on the complex plane are denoted by $(x_i,t_i)$ and the coordinates on the cylinder are denoted by $(\xi_i, \rho_i)$. The transformation of a $GCFT_{1+1}$ primary field $\Phi(x,t)$ under this map is given as\cite{Malvimat:2018izs, Bagchi:2013qva}
\begin{equation}\label{Trans-primary-size}
\tilde{\Phi}(\xi,\rho)=\left(\frac{L}{2\pi i}\right)^{-h_L} e^{\frac{2\pi i}{L}(\xi h_L+\rho h_M)}\Phi(x,t).
\end{equation}
We may also obtain the non-relativistic cross-ratios on the cylinder by using \cref{Trans-size-gcft} as 
\begin{subequations}\label{modified cross}
	\begin{equation}
	\tilde{T} = \frac{\sin\left(\frac{\pi \xi_{12}}{L}\right)\sin\left(\frac{\pi \xi_{34}}{L}\right)}{\sin\left(\frac{\pi \xi_{13}}{L}\right)\sin\left(\frac{\pi \xi_{24}}{L}\right)}, 
	\end{equation}
	\begin{equation}
	\frac{\tilde{X}}{\tilde{T}} = \frac{\pi \rho_{12} }{L}\cot\left(\frac{\pi \xi_{12}}{L}\right)+\frac{\pi \rho_{34} }{L} \cot\left(\frac{\pi \xi_{34}}{L}\right)-\frac{\pi \rho_{13} }{L} \cot\left(\frac{\pi \xi_{13}}{L}\right)-\frac{\pi \rho_{24} }{L} \cot\left(\frac{\pi \xi_{24}}{L}\right).
	\end{equation}
\end{subequations}
Now utilizing \cref{Trans-size-gcft,modified cross} in \cref{Dis. result(GCFT)} we may obtain the OEE for the bipartite mixed state configuration of two disjoint intervals under consideration as  
\begin{equation}\label{dis. finte size result}
\begin{split}
S_o(A_{1}:A_{2})= & \frac{C_L}{6} \log\left[\left(\frac{L}{\pi a}\right)^2 \sin\left(\frac{\pi \xi_{14}}{L}\right)\sin\left(\frac{\pi \xi_{23}}{L}\right)\right]+ \frac{C_L}{12} \log\left(\frac{1+\sqrt{\tilde{T}}}{1-\sqrt{\tilde{T}}}\right) \\
&   +\frac{C_M}{6 }\left[\frac{\pi \rho_{14}}{L}\cot\left(\frac{\pi \xi_{14}}{L}\right)+\frac{\pi \rho_{23}}{L} \cot\left(\frac{\pi \xi_{23}}{L}\right)\right]+\frac{C_M}{12} \frac{\tilde{X}}{\sqrt{\tilde{T}}} \left(\frac{1}{1-\tilde{T}}\right)+\ldots,
\end{split}
\end{equation}
where again the first and the third term denote the EE $S(A_1 \cup A_2)$ for the mixed state $A_1 \cup A_2$. This allows to rewrite the above expression as follows
\begin{equation}
S_o(A_{1}:A_{2})-S{(A_1 \cup A_2)}= \frac{C_L}{12} \log\left(\frac{1+\sqrt{\tilde{T}}}{1-\sqrt{\tilde{T}}}\right)+ \frac{C_M}{12} \frac{\tilde{X}}{\sqrt{\tilde{T}}}\left(\frac{1}{1-\tilde{T}}\right)+\ldots.
\end{equation}
We note here that as earlier, the right-hand-side of the above expression matches exactly with the corresponding EWCS \cite{Basu:2021awn} in the context of flat holography. This is once again consistent with the duality \eqref{duality} and provides strong substantiation for our computations.

\subsubsection{Two disjoint intervals at a finite temperature}
Next we proceed to the computation of the OEE for two disjoint intervals in a $GCFT_{1+1}$ described on a thermal cylinder with circumference equal to the inverse temperature $\beta = 1/T$. We again employ the following conformal map to transform to the thermal cylinder \cite{Bagchi:2013qva}
\begin{equation}\label{Trans-Temp-gcft}
t_{i}=e^{\frac{2\pi\xi_{i}}{\beta}}, \,\,\,\, x_{i}=\frac{2\pi\rho_{i}}{\beta}e^{\frac{2\pi\xi_{i}}{\beta}},
\end{equation}
where $(x_i, t_i)$ denotes the coordinates on the complex plane and $(\xi_i, \rho_i)$ denotes the coordinates on the thermal cylinder. Again the $GCFT$ primaries transform under the above conformal map as follows \cite{Bagchi:2013qva}
\begin{equation}\label{Trans-primary-Temp}
\tilde{\Phi}(\xi,\rho)=\left(\frac{\beta}{2\pi}\right)^{-h_L^{(1)}} e^{\frac{2\pi}{\beta}(\xi h_L^{(1)}+\rho h_M^{(1)})}\Phi(x,t),
\end{equation}
and the $GCFT$ cross-ratios get modified as follows
\begin{subequations}\label{Modified}
	\begin{equation}	
	\hat{T}= \frac{\sinh\left(\frac{\pi \xi_{12}}{\beta}\right)\sinh\left(\frac{\pi \xi_{34}}{\beta}\right)}{\sinh\left(\frac{\pi \xi_{13}}{\beta}\right)\sinh\left(\frac{\pi \xi_{24}}{\beta}\right)},
	\end{equation}
	\begin{equation}  \frac{\hat{X}}{\hat{T}} = \frac{\pi \rho_{12}}{\beta} \coth\left(\frac{\pi \xi_{12}}{\beta}\right)+\frac{\pi \rho_{34}}{\beta} \coth\left(\frac{\pi \xi_{34}}{\beta}\right)-\frac{\pi \rho_{13}}{\beta} \coth\left(\frac{\pi \xi_{13}}{\beta}\right)-\frac{\pi \rho_{24}}{\beta} \coth\left(\frac{\pi \xi_{24}}{\beta}\right).
	\end{equation}
\end{subequations}
Utilizing \cref{Trans-Temp-gcft,Modified} in \cref{Dis. result(GCFT)} we may now obtain the OEE for the bipartite configuration in question to be
\begin{equation}\label{Dis-temp.(GCFT)}
\begin{split}
S_o(A_{1}:A_{2}) = & \frac{C_L}{6} \log\left[\left(\frac{\beta}{\pi a}\right)^2 \sinh\left(\frac{\pi \xi_{14}}{\beta}\right)\sinh\left(\frac{\pi \xi_{23}}{\beta}\right)\right]+ \frac{C_L}{12} \log\left(\frac{1+\sqrt{\hat{T}}}{1-\sqrt{\hat{T}}}\right) \\
&   +\frac{C_M}{6}\left[\frac{\pi \rho_{14}}{\beta}\coth\left(\frac{\pi \xi_{14}}{\beta}\right)+\frac{\pi \rho_{23}}{\beta}\coth\left(\frac{\pi \xi_{23}}{\beta}\right)\right]+\frac{C_M}{12} \frac{\hat{X}}{\sqrt{\hat{T}}} \left(\frac{1}{1-\hat{T}}\right)+\ldots.
\end{split}
\end{equation}
Similar to the previous subsections the first and the last term in the above equation denotes the EE of the mixed state $A_1 \cup A_2$ and we may express the above equation as
\begin{equation}\label{dis finite temp. result}
S_o(A_{1}:A_{2})-S{(A_1 \cup A_2)}= \frac{C_L}{12} \log\left(\frac{1+\sqrt{\hat{T}}}{1-\sqrt{\hat{T}}}\right)+ \frac{C_M}{12} \frac{\hat{X}}{\sqrt{\hat{T}}} \left(\frac{1}{1-\hat{T}}\right)+\ldots.
\end{equation}
We again note that the quantity on the right-hand-side of the above expression matches exactly with the corresponding EWCS \cite{Basu:2021awn} which is in accordance with the duality \cref{duality} in the context of flat holography. This once again provides a consistency check for our construction.

\subsection{OEE for adjacent intervals}
Having computed the OEE for bipartite configurations involving two disjoint intervals, we now turn our attention to the mixed state configurations of two adjacent intervals in $GCFT_{1+1}$s.

\subsubsection{Adjacent intervals at zero temperature}
For this case, we consider two adjacent intervals in a $GCFT_{1+1}$ in its ground state. In particular we consider the two adjacent intervals to be described as $A_1 \equiv [u_1,u_2]$, $A_2 \equiv [u_2,v_2]$ in $GCFT_{1+1}$ which may be obtained by taking the limit $v_1 \to u_2$ in disjoint intervals configuration considered in subsection \ref{sec:disj-gcft}. In this limit, \cref{trace} reduces to the following three point twist correlator,
\begin{equation}\label{gcft-3-point-twist}
\textrm{Tr}(\rho_{A_{1}A_{2}}^{T_{A_{2}}})^{n_{o}}=\left<{\Phi}_{n_{o}}(u_1)\Phi^2_{-n_{o}}(u_2)\Phi_{n_{o}}(v_2)\right> .
\end{equation}
\begin{figure}[H]
	\centering
	\includegraphics[scale=.7]{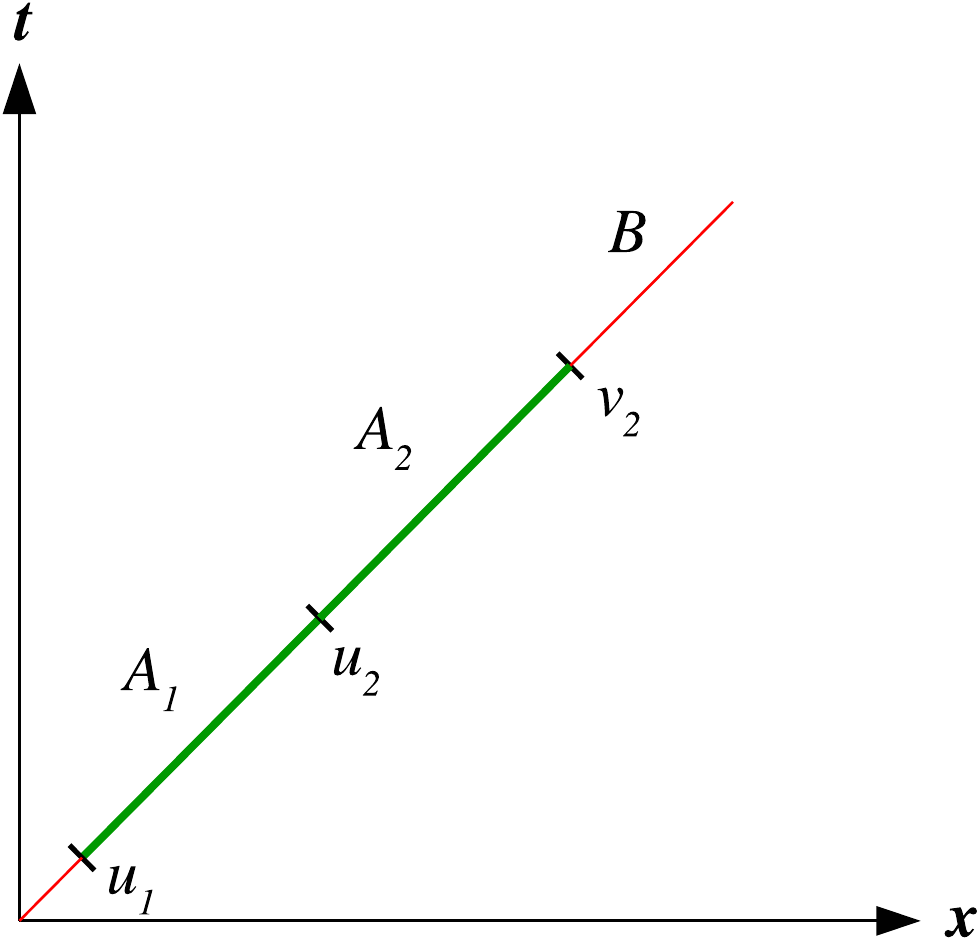}\\
	\caption{Two adjacent intervals in a  $GCFT_{1+1}$ plane.}
	\label{fig3}
\end{figure}
Using the usual form of a $GCFT_{1+1}$ three point correlator in \cref{gcft-3-point} for \cref{gcft-3-point-twist}, we may obtain the OEE for the mixed state of two adjacent intervals under consideration as 
\begin{equation} \label{adj-zero}
S_o(A_{1}:A_{2})= \frac{C_L}{12}\log\left(\frac{t_{12}t_{23}t_{13}}{a^3}\right) +\frac{C_M}{12}\left(\frac{x_{12}}{t_{12}}+\frac{x_{23}}{t_{23}}+\frac{x_{13}}{t_{13}}\right) + \ldots,
\end{equation}
where $a$ is the UV cut-off. The above expression may also be rewritten as 
\begin{equation}\label{adj-zero GCFT}
S_o(A_{1}:A_{2})-S{(A_1 \cup A_2)}= \frac{C_L}{12}\log\left(\frac{t_{12}t_{23}}{a t_{13}}\right) +\frac{C_M}{12}\left(\frac{x_{12}}{t_{12}}+\frac{x_{23}}{t_{23}}-\frac{x_{13}}{t_{13}}\right) + \ldots,
\end{equation}
where we have subtracted the EE for the state $A_1 \cup A_2$ which is given as
\begin{equation}
S{(A_1 \cup A_2)} = \frac{C_L}{6} \log( \frac{t_{13}}{a} ) + \frac{C_M}{6} \left( \frac{x_{13}}{t_{13}} \right).
\end{equation}
Note that we may obtain the result in \cref{adj-zero} by taking an appropriate adjacent limit $(x_2,t_2) \to (x_3,t_3)$ of the corresponding disjoint intervals result in \cref{Dis. result(GCFT)}. We also observe that the expression for the difference of the OEE and the entropy in \cref{adj-zero GCFT} matches with the corresponding EWCS in \cite{Basu:2021awn} apart from an additive constant which is contained in the undetermined OPE coefficient of the corresponding three point twist correlator in \cref{gcft-3-point-twist}. This is in accordance with the duality in \cref{duality}. These observations serve as consistency checks for our results.

\subsubsection{Adjacent intervals in a finite size system}
For this case, we consider the bipartite mixed state configuration of two adjacent intervals described by $A_1 \equiv [(\xi_1, \rho_1), (\xi_2, \rho_2)]$ and $A_2 \equiv [(\xi_2, \rho_2), (\xi_3, \rho_3)]$ in a finite sized $GCFT_{1+1}$ described on a cylinder of circumference $L$. It is again necessary here to compute the corresponding three point twist correlator on the cylinder. To this end, we employ the conformal map in \cref{Trans-size-gcft} and the transformation of the $GCFT_{1+1}$ primaries given in  \cref{Trans-primary-size} to obtain
\begin{equation}\label{Adj. Trans. Size}
\begin{aligned}
\left<\Phi_{n_o}(\xi_1,\rho_1)\Phi^2_{-n_o}(\xi_2,\rho_2)\Phi_{n_o}(\xi_3,\rho_3)\right>
=&\left(\frac{L}{2\pi i}\right)^{-2h_{L}^{(1)}-h_{L}^{(2)}} 
\exp\bigg[\frac{2\pi i}{L}\bigg(\xi_1h_{L}^{(1)}+\xi_2h_{L}^{(2)}+\xi_3h_{L}^{(1)}\\
&+\rho_1h_{M}^{(1)}+\rho_2h_{M}^{(2)}+\rho_3h_M^{(1)}\bigg)\bigg]\,
\left<{\Phi}_{n_{o}}(u_1)\Phi^2_{-n_{o}}(u_2)\Phi_{n_{o}}(v_2)\right>,
\end{aligned}
\end{equation}	
where the three point twist correlator on the right-hand-side is defined on the $GCFT$ complex plane. Utilizing \cref{Adj. Trans. Size,gcft-3-point} in \cref{odd-renyi,oee-def} we may now obtain the OEE for the mixed state under consideration as follows
\begin{equation}\label{adj-L}
\begin{aligned}
S_o(A_{1}:A_{2}) = & \frac{C_L}{12} \log\left[\left(\frac{L}{\pi a}\right)^3 \sin\left(\frac{\pi \xi_{12}}{L}\right)\sin\left(\frac{\pi \xi_{23}}{L}\right)\sin\left(\frac{\pi \xi_{13}}{L}\right)\right] \\
&  +\frac{C_M}{12}\left[\frac{\pi \rho_{12}}{L}\cot\left(\frac{\pi \xi_{12}}{L}\right)+\frac{\pi \rho_{23}}{L}\cot\left(\frac{\pi \xi_{23}}{L}\right)+\frac{\pi \rho_{13}}{L}\cot\left(\frac{\pi \xi_{13}}{L}\right)\right]+\ldots.
\end{aligned}
\end{equation}
It is instructive to rewrite the above result as
\begin{equation}\label{adj-L-subtracted}
\begin{split}
S_o(A_{1}:A_{2})-S{(A_1 \cup A_2)} = \frac{C_L}{12} \log\left[\frac{L}{\pi a} \frac{\sin\left(\frac{\pi \xi_{12}}{L}\right)\sin\left(\frac{\pi \xi_{23}}{L}\right)}{\sin\left(\frac{\pi \xi_{13}}{L}\right)}\right]& +\frac{C_M}{12 }\Bigg[\frac{\pi \rho_{12}}{L}\cot\left(\frac{\pi \xi_{12}}{L}\right)\\+\frac{\pi \rho_{23}}{L} \cot\left(\frac{\pi \xi_{23}}{L}\right)
&-\frac{\pi \rho_{13}}{L} \cot\left(\frac{\pi \xi_{13}}{L}\right)\Bigg]+\ldots,
\end{split}
\end{equation}
where the EE $S(A_1 \cup A_2)$ for the mixed state $A_1 \cup A_2$ is given as
\begin{equation}
S{(A_1 \cup A_2)}=\frac{C_L}{6} \log\left[\frac{L}{\pi a} \sin(\frac{\pi \xi_{13}}{L})\right]+\frac{C_M}{6}\frac{\pi \rho_{13}}{L}\cot\left(\frac{\pi \xi_{13}}{L}\right)+ \ldots.
\end{equation}
Similar to the previous case, by the application of an appropriate adjacent limit in the disjoint intervals result given in \cref{dis. finte size result} we may reproduce the above adjacent intervals result in \cref{adj-L}. Also note that the right-hand-side of \cref{adj-L-subtracted} matches with the corresponding EWCS obtained in \cite{Basu:2021awn} up to an additive constant which is contained in the undetermined OPE coefficient of the corresponding three point twist correlator.

\subsubsection{Adjacent intervals at a finite temperature} \label{sec:adj-T-GCFT}
We now proceed to the case of two adjacent intervals at a finite temperature $T$ in a $GCFT_{1+1}$. To this end, we consider the bipartite mixed state described by $A_1 \equiv [(\xi_1, \rho_1), (\xi_2, \rho_2)]$ and $A_2 \equiv [(\xi_2, \rho_2), (\xi_3, \rho_3)]$ in a $GCFT_{1+1}$ defined on a thermal cylinder with the circumference given by the inverse temperature $\beta = 1/T$. Similar to the previous subsection, we need to compute the three point twist correlator in \cref{gcft-3-point-twist} on the thermal cylinder. This is done by utilizing the map \eqref{Trans-Temp-gcft} and the transformation of $GCFT$ primaries in \cref{Trans-primary-Temp} to obtain 
\begin{equation}\label{ft}
\begin{aligned}
\left<\Phi_{n_o}(\xi_1,\rho_1)\Phi^2_{-n_o}(\xi_2,\rho_2)\Phi_{n_o}(\xi_3,\rho_3)\right>_{\beta}
&=\left(\frac{\beta}{2\pi }\right)^{-2h_{L}^{(1)}-h_{L}^{(2)}} 
\exp\Bigg[\frac{2\pi}{\beta}(\xi_1h_{L}^{(1)}+\xi_2h_{L}^{(2)}+\xi_3h_{L}^{(1)}\\
&+\rho_1h_{M}^{(1)}+\rho_2h_{M}^{(2)}+\rho_3h_M^{(1)})\Bigg]\,
\left<{\Phi}_{n_{o}}(u_1)\Phi^2_{-n_{o}}(u_2)\Phi_{n_{o}}(v_2)\right>.
\end{aligned}
\end{equation}
Here the subscript $\beta$ on the left denotes that the correlator is described on the thermal cylinder and the three point twist correlator on the right-hand-side is described on the $GCFT$ complex plane. Now utilizing \cref{ft,gcft-3-point} we may obtain the OEE for the mixed states of two adjacent intervals at a finite temperature to be
\begin{equation}\label{adj-T}
\begin{aligned}
S_o(A_{1}:A_{2})= &  \frac{C_L}{12} \log\left[\left(\frac{\beta}{\pi a}\right)^3 \sinh\left(\frac{\pi \xi_{12}}{\beta}\right)\sinh\left(\frac{\pi \xi_{23}}{\beta}\right)\sinh\left(\frac{\pi \xi_{13}}{\beta}\right)\right] \\
&   +\frac{C_M}{12}\left[\frac{\pi \rho_{12}}{\beta}\coth\left(\frac{\pi \xi_{12}}{\beta}\right)   +\frac{\pi \rho_{23}}{\beta}\coth\left(\frac{\pi \xi_{23}}{\beta}\right)+\frac{\pi \rho_{13}}{\beta}\coth\left(\frac{\pi \xi_{13}}{\beta}\right)\right]+\ldots,
\end{aligned}
\end{equation}
where $a$ is a UV cut off of the $GCFT_{1+1}$. We may again rearrange the above expression to obtain 
\begin{equation}\label{adj-T-subtracted}
\begin{split}
S_o(A_{1}:A_{2})-S{(A_1 \cup A_2)} = \frac{C_L}{12} \log\Bigg[\frac{\beta}{\pi a}& \frac{\sinh\left(\frac{\pi \xi_{12}}{\beta}\right)\sinh\left(\frac{\pi\xi_{23}}{\beta}\right)}{\sinh\left(\frac{\pi \xi_{13}}{\beta}\right)}\Bigg]   +\frac{C_M}{12}\Bigg[\frac{\pi \rho_{12}}{\beta} \coth\left(\frac{\pi \xi_{12}}{\beta}\right) \\ & +\frac{\pi \rho_{23}}{\beta}\coth\left(\frac{\pi \xi_{23}}{\beta}\right)
-\frac{\pi \rho_{13}}{\beta} \coth\left(\frac{\pi \xi_{13}}{\beta}\right)\Bigg]+\ldots,
\end{split}
\end{equation}	
where the EE $S(A_1 \cup A_2)$ for the subsystem $A_1 \cup A_2$ is given as
\begin{equation}
S{(A_1 \cup A_2)}=\frac{C_L}{6} \log\left[\frac{\beta}{\pi a} \sinh(\frac{\pi \xi_{13}}{\beta})\right]+\frac{C_M}{6}\frac{\pi \rho_{13}}{\beta}\coth\left(\frac{\pi \xi_{13}}{\beta}\right)+ \ldots.
\end{equation}
Again we note that through the appropriate adjacent limit in the corresponding disjoint intervals result in \cref{dis finite temp. result} we may obtain the result for the OEE for the mixed state configuration of two adjacent intervals in \cref{adj-T}. We also observe that the right-hand-side of \cref{adj-T-subtracted} matches with the corresponding EWCS computed in the context of flat space holography in \cite{Basu:2021awn} apart from an additive constant which is contained in the undetermined OPE coefficient of the corresponding three point twist correlator. These serve as consistency checks for our computations.

\subsection{OEE for a single interval}
Having computed the OEE for the mixed states of two disjoint and two adjacent intervals, finally, in this subsection we proceed to the computation of the OEE for bipartite pure and mixed state configurations involving a single interval in $GCFT_{1+1}$s.

\subsubsection{Single interval at zero temperature}
For this case, we consider a single boosted interval $A_1=[(x_1,t_1),(x_2,t_2)]$ at zero temperature which describes a bipartite pure state in a $GCFT_{1+1}$. We may obtain this configuration through the limit $u_2 \to v_1$ and $v_2 \to u_1$ in the disjoint intervals construction described in subsection \ref{sec:disj-gcft}. Here the interval $A \equiv A_1 \cup A_2$ describes the full system with $B$ as a null set and consequently the state described by the density matrix $\rho_A$ is a pure state. In this limit the four point twist correlator in \cref{trace} reduces to the following two point twist correlator,
\begin{equation}\label{single-two-point-gcft}
\textrm{Tr}(\rho_{A}^{T_{A_{2}}})^{n_{o}}=\left<\Phi_{n_{o}}^{2}(u_{1})\Phi_{-n_{o}}^{2}(v_{1})\right>,
\end{equation}
where the twist operators $\Phi_{n_{o}}^{2}$ and $\Phi_{-n_{o}}^{2}$ have the weights $h_{L}^{(2)}=h_{L}^{(1)}$ and $h_{M}^{(2)}=h_{M}^{(1)}$. Now utilizing the usual form of a $GCFT$ two point correlator given in \cref{gcft two point} for the above twist correlator, we may obtain the OEE for the given pure state of a single interval as follows
\begin{equation}\label{single-gcft}
S_o(A_{1}:A_{2})= \frac{C_L}{6} \log\left(\frac{t_{12}}{a}\right)+\frac{C_M}{6} \left(\frac{x_{12}}{t_{12}}\right)+ \ldots ,
\end{equation}
where $a$ is a UV cut-off for the $GCFT_{1+1}$. We observe here that the OEE obtained above matches exactly with the corresponding EWCS \cite{Basu:2021awn} and with the EE \cite{Basu:2015evh} for the single interval $A_1$. This is in conformity with the quantum information theory expectation that for a pure state the OEE should reduce to the EE for the single interval describing the bipartite state \cite{Tamaoka:2018ned}.

\subsubsection{Single interval in a finite size system}
In this subsection, we focus on the computation of the OEE for the pure state of a single interval in a finite sized $GCFT_{1+1}$. To this end, we consider a single interval $A$ in a $GCFT_{1+1}$ defined on cylinder with circumference $L$. Using \cref{Trans-primary-size} in \cref{gcft two point}, we may obtain the corresponding two point twist correlator on this cylinder as follows
\begin{equation}\label{two point on cylinder}
\left<\Phi^2_{n_o}(\xi_1,\rho_1)\Phi^2_{-n_o}(\xi_2,\rho_2)\right>
=\left[\frac{L}{\pi}\sin\left(\frac{\pi \xi_{12}}{L}\right)\right]^{-2h_{L}^{(2)}}
\exp\left[-2h_{M}^{(2)}\frac{\pi \rho_{12}}{L}\cot\left(\frac{\pi \xi_{12}}{L}\right)\right],
\end{equation}
where $(\xi_i,\rho_i)$ are the endpoints of the interval $A$ on the cylinder. Now by using \cref{two point on cylinder} and \cref{odd-renyi,oee-def}, we may obtain the OEE for the single interval in a finite sized system as 
\begin{equation}
S_o(A_{1}:A_{2})= \frac{C_L}{6}\log\left(\frac{L}{\pi a}\sin\frac{\pi \xi_{12}} {L}\right)+ \frac{C_M }{6} \frac{\pi \rho_{12}}{L} \cot{\frac{\pi \xi_{12}}{L}}+\ldots.
\end{equation}
We again observe the consistent behaviour of the OEE obtained above to match exactly with the EE and the EWCS for the corresponding single interval \cite{Basu:2021axf}. These matchings serve as a consistency checks for our result.

\subsubsection{Single interval at a finite temperature}
For this final case, we consider a single interval $A \equiv [(-\xi,-\rho),(0,0)]$ in a $GCFT_{1+1}$ defined on a thermal cylinder whose circumference is equal to the inverse temperature $\beta$. Similar to the relativistic case discussed in subsection \ref{sec:single-T-CFT}, it is necessary to consider the single interval in question to be sandwiched between two large but finite auxiliary intervals $B_1 \equiv [(-L,-y),(-\xi,-\rho)]$ and $B_2 \equiv [(0,0),(L,y)]$ adjacent on either side\footnote{In \cite{Basak:2022cjs, Malvimat:2018izs}, the authors found that a similar construction was required to appropriately compute the entanglement negativity and the reflected entropy for the same configuration of a single interval at a finite temperature in a $GCFT_{1+1}$.}. The OEE is then computed with finite auxiliary intervals and finally the bipartite limit described $B \equiv B_1 \cup B_2 \to A^c$ or $L \to \infty$ is taken to restore the original configuration.

With the presence of two auxiliary intervals, the OEE may then be obtained by computing the following four twist correlator on the thermal cylinder,
\begin{equation}\label{single-finite-T(GCFT)}
S_o(A:B)= \lim_{L\to \infty} \lim_{n_{o}\to 1}\frac{1}{1-n_o}
\ln\left[\left<\Phi_{n_o}(-L,-y)\,\Phi^2_{-n_o}(-\xi,-\rho)\,\Phi^2_{n_o} (0,0)\,\Phi_{-n_o}(L,y)\right>_\beta\right].
\end{equation}
The above four point twist correlator on a $GCFT$ complex plane is given by \cite{Malvimat:2018izs}
\begin{equation} \label{Four pt.(II)(GCFT)}
\begin{aligned}
\left<\Phi_{n_o}(x_1,t_1)\,\Phi^2_{-n_o}(x_2,t_2)\,\Phi_{n_o}^2(x_3,t_3)\,\Phi_{-n_o}(x_4,t_4)\right>
=&\frac{\tilde{k}_{n_o}}{t_{14}^    
	{2h_L}\,t_{23}^{2h_L}}
\frac{\mathcal{F}_{n_o}\left(T,\frac{X}{T}\right)}{T^{h_L
}}\\
& \times \exp \bigg [ - 2 h_M
\frac{x_{14}}{t_{14}}
- 2 h_M \frac{x_{23}}{t_{23}} - h_M \frac{X}{T} \bigg ],
\end{aligned}
\end{equation}
where $\tilde{k}_{n_o}$ is a constant, $X$ and $T$ are the $GCFT_{1+1}$ cross-ratios given in \cref{GCFT cross ratio}, $h_L \equiv h_L^{(1)} = h_L^{(2)}$, $h_M \equiv h_M^{(1)} = h_M^{(2)}$ and $\mathcal{F}_{n_o}$ is a non-universal function of the cross-ratios and depend on the full operator content of the theory. In the limits $T \to 1$ and $T \to 0$, the non-universal function $\mathcal{F}_{n_o}$ have the following behaviour \cite{Malvimat:2018izs}
\begin{equation}
\mathcal{F}_{n_o}(1,0)=1, \hspace{2cm} \mathcal{F}_{n_o}\left(0,\frac{X}{T}\right)= C_{n_{o}},
\end{equation} 
where $C_{n_{o}}$ is a constant that depends on the full operator content of the theory.

We may utilize the conformal map \eqref{Trans-Temp-gcft} and transformation of the $GCFT$ primaries given in \cref{Trans-primary-Temp} to obtain the required four point twist correlator in \cref{single-finite-T(GCFT)} on the thermal cylinder as
\begin{equation}
\begin{aligned}
&\left<\Phi_{n_o}(-L,-y)\,\Phi^2_{-n_o}(-\xi,-\rho)\,\Phi^2_{n_o} (0,0)\,\Phi_{-n_o}(L,y)\right>_\beta\\
&\qquad  \qquad \qquad = \frac{k_{n_o} k_{n_o / 2}^2}
{T^{h_L}}\left [\frac{\beta}
{\pi}\sinh\left( \frac{2\pi L}{\beta}\right )\right ]^{-2h_L}
\left [\frac{\beta}{\pi}\sinh \left(\frac{\pi\xi}{\beta}\right)\right ]^{-2h_L}\\
&\qquad \qquad \qquad ~~ \times \exp\left[-\frac{2\pi y}{\beta}\coth\left(\frac{2\pi L}{\beta}\right)2h_M-\frac{2\pi \rho}{\beta}\coth\left(\frac{\pi\xi}{\beta}\right)2h_M-\frac{X}{T}
h_M\right]\mathcal{F}_{n_o}\left(T, \frac{X}{T}\right).
\end{aligned}
\end{equation}
Now using the above expression in \cref{single-finite-T(GCFT)} and taking the bipartite limit $L \to \infty$ subsequent to the replica limit $n_o \to 1$, we may obtain the OEE for the mixed state under consideration to be
\begin{equation}\label{sing-T-gcft}
\begin{aligned}
S_o(A:B) = &   \lim_{L\to \infty}\left[\frac{C_L}{6}\log\left(\frac{\beta}{\pi a} \sinh\frac{2\pi L}{\beta}\right) + \frac{C_M}{6}\frac{\pi (2y) }{\beta}\coth\frac{2\pi L}{\beta}\right] +  \frac{C_L}{6} \log\left(\frac{\beta}{\pi a}\sinh\frac{\pi \xi}{\beta}\right)\\
&  +\frac{C_M}{6} \frac{\pi \rho}{\beta}\coth\frac{\pi \xi}{\beta}-\frac{C_L}{6}\frac{\pi \xi }{\beta}-\frac{C_M}{6}\frac{\pi\rho }{\beta}+ f\left(e^{-\frac{2\pi\xi}{\beta}},-\frac{2\pi\rho}{\beta}\right)+ \ldots ,
\end{aligned}
\end{equation}
where $a$ is a UV cut off and the non universal function $f (T,X/T)$ is defined as \cite{Malvimat:2018izs}
\begin{equation}
f\left(T,\frac{X}{T}\right) \equiv \lim_{n_o \to 1} \ln \left[ \mathcal{F}_{n_o} \left(T, \frac{X}{T} \right) \right].
\end{equation}
Note that the divergent first term inside the parenthesis in \cref{sing-T-gcft} describes the EE of the total thermal system $A \cup B_1 \cup B_2$ with the bipartite limit $B = B_1 \cup B_2 \to A^c$. We may extract the finite part of the OEE for the mixed state of the single interval under consideration as
\begin{equation} \label{sing-T-subtracted}
\begin{aligned}
S_o(A:B)-S{(A \cup B)}=\frac{C_L}{6} \log&\left(\frac{\beta}{\pi a}\sinh\frac{\pi \xi}{\beta}\right)+\frac{C_M}{6} \frac{\pi \rho}{\beta}\coth\frac{\pi \xi}{\beta}\\& -\frac{\pi \xi C_{L}}{6\beta}-\frac{\pi\rho C_{M}}{6\beta} +f\left(e^{-\frac{2\pi\xi}{\beta}},-\frac{2\pi\rho}{\beta}\right)+ \ldots,
\end{aligned}
\end{equation}
where the divergent total entropy $S(A \cup B)$ has been subtracted. We observe that the above result may be expressed in a more instructive way as follows
\begin{equation}
S_o(A:B)-S{(A \cup B)} = S(A) - S^\text{th}(A) + f \left( e^{ - \frac{2 \pi \xi}{\beta}} , - \frac{2 \pi \rho}{\beta} \right) + \ldots,
\end{equation}
where $S(A)$ denotes the EE for the single interval $A$ \cite{Bagchi:2014iea} and $S^\text{th}(A)$ denotes the thermal contribution to the EE. The above expression highlights that the universal part of the OEE does not include contributions from the thermal correlations. We note here that the right-hand-side of \cref{sing-T-subtracted} matches with the upper bound of the corresponding EWCS obtained in \cite{Basu:2021awn} apart from an additive constant. This extra additive constant in the EWCS is contained in the non-universal function $f (T, X/T)$ and may be obtained through a proper large central charge monodromy analysis of the four point twist correlator in \cref{single-finite-T(GCFT)}.

\section{Summary and conclusions}\label{sec4}
To summarize, in this article we have obtained the odd entanglement entropy for bipartite states in of $(1+1)$-dimensional holographic relativistic and non-relativistic (Galilean) conformal field theories through appropriate replica techniques. From our results we have verified the proposed duality of the bulk EWCS with the difference between the OEE and the EE for bipartite states in holographic $CFT_{1+1}$ and $GCFT_{1+1}$. In this context we have demonstrated the extension of the above duality for $GCFT_{1+1}$s dual to bulk $(2+1)$-dimensional asymptotically flat geometries.

For the relativistic case  we have obtained the OEE for bipartite pure and mixed states at zero and finite temperature and for finite sized system involving two disjoint, two adjacent and a single interval in $CFT_{1+1}$s through a replica technique. In this context we have obtained the corresponding results for the various bipartite states in relativistic $CFT_{1+1}$  which were missing in the literature. Furthermore we have also verified  the holographic duality of the difference between the OEE and the EE with the corresponding bulk EWCS for the above mentioned bipartite states. We also found that our result for the adjacent intervals at zero temperature matches with earlier works in the literature in the context of 1-dimensional harmonic spin chains and for gravitational path integral computations based on fixed area states. These serve as consistency checks for our computations.

Subsequent to the above we have investigated non-relativistic holographic $GCFT_{1+1}$s and established an appropriate replica technique to compute the OEE for bipartite states. In this context have computed the OEE for such bipartite states in $GCFT_{1+1}$s involving two disjoint, two adjacent and a single interval at zero and finite temperatures and for finite sized systems. Furthermore we have also compared our results with the corresponding bulk EWCS to verify and extend its duality with the difference between the OEE and EE to a flat holographic scenario.

In the above connection we have obtained the OEE for the mixed state configuration of two disjoint intervals at zero and a finite temperature and in a finite sized system utilizing a geometric monodromy technique to obtain the large central charge limit of the corresponding four point twist field correlator in the $GCFT_{1+1}$. Interestingly for the all the cases we observed that the difference between the OEE and the EE are exactly equal to the corresponding bulk EWCS computed earlier in the literature, which substantiates our computations and extends the duality to the framework of flat holography.

Following this we have obtained the OEE for the mixed state configuration of two adjacent intervals at zero and a finite temperature and in a finite sized system in $GCFT_{1+1}$s. As a consistency check we have also obtained the above results from the OEE for disjoint intervals through a suitable adjacent limit. Once again we have compared our results with bulk EWCS computed earlier in the literature in the context of flat holography and observed the matching with the functional form of the difference between the OEE and the EE. However we should mention here that the bulk EWCS for these cases involve an extra additive constant arising from the undetermined OPE coefficient of the corresponding three point twist correlator in the dual  $GCFT_{1+1}$.

Subsequently, we consider the pure state configuration of a single interval at zero temperature and in a finite sized system. We have found that the OEE for these pure state configurations exactly match with the corresponding entanglement entropies (EE) as dictated by the quantum information theory. Finally for the mixed state of a single interval at a finite temperature in the $GCFT_{1+1}$s it was required to utilize a construction involving two large but finite auxiliary intervals adjacent on either sides of the single interval to correctly compute the corresponding OEE. We mention here that similar constructions have been employed in the  literature for the computation of entanglement negativity and reflected entropy for this specific mixed state configuration in both relativistic and non-relativistic scenarios. Furthermore it was observed the OEE for this bipartite state involves a divergent part due the EE of the total (infinite) system. However, on subtraction of this divergent contribution from the universal part of the OEE, we observe that the finite part matches with the upper bound of the corresponding bulk EWCS computed earlier for flat space holography. Additionally, in the appendix \ref{appendix_A} we have also reproduced the expression for the OEE for two disjoint intervals in a $GCFT_{1+1}$ through an appropriate non-relativistic limit of the corresponding result in the usual relativistic $CFT_{1+1}$. All of these detailed comparisons and matches serve as strong consistency checks for our computations. 

In conclusion we state that the characterization of entanglement for bipartite states in conformal field theories and its relation with space time holography is an extremely rich field for further investigation and has provided significant insights into diverse issues in condensed matter theories and also in quantum gravity and black hole information. There are exciting open avenues in this context for further investigation of various other entanglement measures defined in quantum information theory which is expected to provide further elucidation of crucial issues in the above disciplines. We hope to return to these fascinating issues in the near future.

\appendices
\section{Limiting Analysis} \label{appendix_A}
In this appendix we perform a limiting analysis and show that the OEE for a bipartite mixed state in a $GCFT_{1+1}$ computed in \cref{sec3} can be obtained from the corresponding relativistic result in \cref{sec2} through an appropriate non-relativistic limit. In this regard, the parametric \.In\"on\"u-Wigner contraction given in \cref{Inonu-Wigner} may be expressed in terms of the coordinates of the $CFT_{1+1}$ complex plane as
\begin{equation}\label{Contractions}
	z \to t + \epsilon x \, \, , \qquad \bar{z} \to t - \epsilon x \, ,
\end{equation}
where $\epsilon \to 0$. The central charges of the $GCA_{1+1}$ may also be related to those of the parent relativistic theory as \cite{Bagchi:2009pe, Basu:2021awn}
\begin{equation}\label{gcabms}
	C_L = c + \bar{c} \quad, \quad C_M = \epsilon(c - \bar{c}) \, .
\end{equation}
For unequal central charges $c$ and $\bar c$ for the holomorphic and anti-holomorphic sectors, the OEE for the bipartite mixed state of two disjoint intervals in a $CFT_{1+1}$ given in \cref{Dis.} may be expressed as
\begin{equation}\label{Dis- CFT(1)}
S_o(A_{1}:A_{2})= S(A_{1}\cup A_{2})+\frac{c}{12}\log \left[\frac{1+\sqrt{x}}{1-\sqrt{x}}\right]+\frac{\bar{c}}{12} \log \left[ \frac{1 + \sqrt{\bar{x}}}{1 - \sqrt{\bar{x}}} \right]+\ldots ,
\end{equation}
where the entanglement entropy $S (A_1 \cup A_2)$ is similarly expressed in terms of the unequal central charges $c$ and $\bar c$. Utilizing \cref{Contractions}, we may write the $CFT_{1+1}$ cross ratios $x, \bar x$ in terms of the $GCFT_{1+1}$ cross ratios $X, T$ as
\begin{equation}\label{CrossTrans}
x\rightarrow T\left(1+\epsilon \frac{X}{T}\right)\,\,,\qquad \bar{x}\rightarrow T\left(1-\epsilon \frac{X}{T}\right)\,.
\end{equation}
We may now obtain the OEE for the corresponding bipartite configuration in the $GCFT_{1+1}$ up to linear order in $\epsilon$ by utilizing \cref{CrossTrans,Contractions} in \cref{Dis- CFT(1)} to be
\begin{equation}
\begin{aligned}
S_o(A_{1}:A_{2})  =&\frac{C_L}{6} \log\left(\frac{t_{14}t_{23}}{a^2}\right)+ \frac{C_L}{12} \log\left(\frac{1+\sqrt{T}}{1-\sqrt{T}}\right)\\
& + \frac{C_M}{6} \left(\frac{x_{14}}{t_{14}}+\frac{x_{23}}{t_{23}}\right)+\frac{C_M}{12} \frac{X}{\sqrt{T}} \left(\frac{1}{1-T}\right)+ \mathcal{O} (\epsilon).
\end{aligned}
\end{equation}
It is remarkable that the above expression matches exactly with the corresponding replica technique result in \cref{Dis. result(GCFT)} up to the leading order which provides a consistency check for our computations. Similarly, we have checked that this limiting behaviour of the OEE in $GCFT_{1+1}$ also holds for the other bipartite configurations discussed in this article as well. We note here that although it is possible to obtain the expression for the OEE for bipartite subsystems in $GCFT_{1+1}$s following the above procedure, however the above limiting analysis lacks the information about the structure of the replica manifold which is important in the study of the R\'enyi generalization of the OEE and its application in holography.

\bibliographystyle{utphys}

\bibliography{reference}

\end{document}